\documentclass[
aps,prd,
floatfix,superscriptaddress,
twocolumn
, notitlepage
]{revtex4-1}
\usepackage{graphicx,subfigure}
\usepackage{amsmath,amssymb,amsfonts}
\usepackage{bm}
\usepackage{color}
\usepackage{comment}
\usepackage{lineno}

\usepackage{soul}
\usepackage{ulem}
\setulcolor{blue}
\setstcolor{red}
\sethlcolor{yellow}

\def \be {\begin{equation} }
\def \ee {\end{equation}}

\def \bem {\begin{multline}}
\def \eem {\end{multline}}

\def \bes {\begin{subequations} }
\def \ees {\end{subequations}}

\def \pd {\partial}

\def \a {\alpha}
\def \b {\beta}

\def \d {\delta}
\def \e {\epsilon}

\def \k {\kappa}

\def \t {\tau}

\def \G {\Gamma}

\def \<{\langle}
\def \>{\rangle}
\def \+{\dagger}
\def \({\left(}
\def \){\right)}
\def \[{\left[}
\def \]{\right]}

\def \tM {\tilde{M}}
\def \tl {\tilde{\lambda}}

\def \ttau {\tilde{\tau}}

\def \tF {\tilde{F}}
\def \tsigma{\tilde{\sigma}}

\def \equ {\text{eq}}

\def \eff{\text{eff}}
\def \rel {\text{rel}}

\def \sgn {\text{sgn}}

\def \trel {\tau_{\rm rel}}

\def \tG {\tilde{G}}

\def \teff {\tau_{\rm eff}}

\def \tch {\tau_{\rm quench}}
\def \tKZ {\tau_{\rm KZ}}

\def \lKZ {l_{\rm KZ}}

\def \tKZ {\tau_{\rm KZ}}
\def \lKZ {l_{\rm KZ}}
\def \thetaKZ {\theta_{\rm KZ}}
\def \xieq {\xi_{\rm eq}}

\def \tqxi {\tau_{\rm quench}^\xi}
\def \tqth {\tau_{\rm quench}^\theta}
\def \teff {\tau_{\rm eff}}

\def \xmin {\xi_{\rm min}}
\def \thKZ {\theta_{\rm KZ}}

\begin{document}

\title{Universal off-equilibrium scaling of critical cumulants in the QCD phase diagram}
\author{Swagato~Mukherjee}
\affiliation{Department of Physics,
Brookhaven National Laboratory, Upton, New York 11973-5000}
\author{Raju~Venugopalan}
\affiliation{Department of Physics,
Brookhaven National Laboratory, Upton, New York 11973-5000}
\affiliation{Institut f\"{u}r Theoretische Physik, Universit\"{a}t Heidelberg, Philosophenweg 16, 69120 Heidelberg, Germany}
\author{Yi~Yin}
\affiliation{Department of Physics,
Brookhaven National Laboratory, Upton, New York 11973-5000}

\date{ \today}


%
\begin{abstract}
Exploiting the universality between the QCD critical point and the three dimensional Ising model, closed form expressions derived~\cite{Mukherjee:2015swa} for non-equilibrium critical cumulants on the crossover side of the critical point reveal that they can differ both in magnitude and sign from equilibrium expectations. We demonstrate here that key elements of the Kibble-Zurek framework of non-equilibrium phase transitions can be employed to describe the dynamics of these critical cumulants. Our results suggest that observables sensitive to critical dynamics in heavy-ion collisions should be expressible as universal scaling functions, thereby providing powerful model independent guidance in searches for the QCD critical point. 

\end{abstract}
\maketitle
Theoretical work on the phase diagram of Quantum Chromodynamics
(QCD)~\cite{Stephanov:2004wx,Stephanov:2007fk,Fukushima:2010bq} in the temperature $T$ and baryon chemical potential $\mu_B$ plane suggests the existence of a critical end point (CEP), the end point of a line of first-order phase transitions, that separates, in the chiral limit, a chirally
symmetric quark-gluon plasma (QGP) phase from a hadron matter phase. This CEP is widely believed to lie in the static universality class of
the three-dimensional Ising model~\cite{Berges:1998rc,Halasz:1998qr}. 
A definitive characterization of the phase diagram is hindered by the
sign problem in lattice QCD at finite $\mu_B$; nevertheless, significant progress has been made in extending lattice
thermodynamics from the finite temperature $T\neq 0, \mu_B=0$ axis into the domain of finite $\mu_B$~\cite{Soltz:2015ula,Ding:2015ona}.

A parallel intensive experimental effort is underway to locate and
characterize this critical point through a beam energy scan
(BES) of heavy ion collisions at the Relativistic Heavy Ion Collider
(RHIC), from the highest center of mass energies ($\sqrt{s}= 200$
GeV/nucleon) down to energies per nucleon a few times the nucleon
mass~\cite{Heinz:2015tua,Asakawa:2015ybt,STAR-wp,Nahrgang:2016ayr}. 
The fireballs created in such collisions traverse trajectories in the
$T$-$\mu_B$ plane as they expand and cool before freezing out in a
shower of hadrons.
If the initial conditions are propitious, their dynamics can be
expressed in terms of ``protocols"--classes of trajectories in the relevant parameter
space with differing sensitivity to critical fluctuations of the universal Ising order parameter. In the BES, protocols on the crossover side of the
CEP are most likely, and we will restrict our attention to these. 

Crossover protocols
are subject to the critical slowing down of the
relaxation rate of critical fluctuations. According to the
theory of dynamical critical phenomena~\cite{RevModPhys.49.435}, 
the relaxation time for critical modes is related to their equilibrium correlation length as $\teff\sim
\xi_{\rm eq}^{z}$, where the dynamic scaling exponent $z=3$ for
QCD~\cite{Son:2004iv,Fujii:2003bz,Fujii:2004za,Fujii:2004jt} lies in
the model H universality class. Recently, employing the Fokker-Planck master equation describing the non-equilibrium dynamics of critical fluctuations~\cite{RevModPhys.49.435}, we derived closed form expressions for the temporal evolution of the first four cumulants $\kappa_{n=1,2,3,4}$ of the zero mode of the critical field ~\cite{Mukherjee:2015swa}. This work significantly extended prior work on Gaussian fluctuations~\cite{Berdnikov:1999ph} and showed that both the magnitude and the sign of off-equilibrium non-Gaussian cumulants could differ from their equilibrium counterparts~\cite{Stephanov:2008qz,Asakawa:2009aj,Friman:2011pf,Stephanov:2011pb}. 

While memory effects persisting from critical slowing down could thus be detectable in the BES analyses, our results were sensitive to a number of non-universal inputs governing the protocols that include i) the mapping of the Ising variables -- the reduced temperature $r = (T-T_c)/T_c$ (with $T_c$ denoting the Ising critical temperature) and the rescaled magnetic field $h$ -- to the QCD thermodynamic variables $T,\mu_B$,  
ii) the details of trajectories in QCD phase diagram and  iii) the relaxation rate of the critical mode $\teff$.
We shall henceforth collectively label these non-universal inputs with the symbol $\Gamma$. Uncertainties in $\G$  can be reduced by careful modeling of the hydrodynamical evolution of the fireball and by further developments in lattice QCD studies at finite $\mu_B$. However our prior results suggest that the model dependence of the critical cumulants will survive. 

In this letter, we will show that significant progress towards model independent results for $\kappa_n$ can be achieved by employing the Kibble-Zurek (KZ) framework of non-equilibrium phase transitions to express the critical cumulants $\kappa_n$ for diverse protocols in terms of universal scaling functions. The KZ framework was initiated by Kibble to describe the formation and evolution of topological defects in cosmological phase transitions~\cite{KIBBLE1980183}. It was generalized to describe critical phenomena in a variety of contexts by Zurek~\cite{1985Natur.317..505Z,Zurek:1996sj}; a fruitful application is in the description of Quantum Phase Transitions~\cite{PhysRevB.72.161201}. 
Experimental observations of KZ scaling in various condensed matter systems have also been reported; for a recent example, see Ref.~\cite{nature13}. 
Our work is inspired by a study of KZ dynamics in terms of the universal scaling of correlation functions~\cite{PhysRevB.86.064304}. For further discussion, employing powerful holographic techniques, see Ref.~\cite{Chesler:2014gya} and references within. 

We begin by noting that a reduction in the number of parameters is seen already for equilibrium Ising critical cumulants which can be expressed as $\kappa^{\rm eq}_{n}\sim \xieq^{-\frac{1}{2}+\frac{5}{2}(n-1)} f^{\rm eq}_{n}(\theta)$, where $\xieq(r,h)$ depends universally on $r,h$ and $\theta$ is related to the product $\(r^{-5/3} h\)$
~\footnote{In a linear parametrization model of the equation of state~\cite{PhysRevLett.23.1098,ZinnJustin:1999bf},
$r^{-5/3} h =  (1-\theta^{2})^{-5/3}\( 3-2\theta^{2}\)$ modulo an overall normalization factor. 
Throughout this work, we will use approximate rational values of critical exponents, 
i.e. $(\a,\b,\gamma,\nu, \delta, \eta)=(0,1/3,4/3,5,0)$, which are within a few percent of their exact values as summarized in Ref.~\cite{ZinnJustin:1999bf}.
}.
To address the possibility of an analogous off-equilibrium scaling, consider a system undergoing a slow quench, where initially $\teff$ of the critical mode is much smaller than the quench times
\begin{equation}
\tqxi = \Big|\frac{\xi_{\rm eq}(\tau)}{\partial_\tau \xi_{\rm eq}(\tau)}\Big| \,\,,
\qquad 
\tqth = \Big|\frac{\theta(\tau)}{\partial_\tau\theta(\tau)}\Big|\, ,
\label{eq:quench-rates}
\end{equation}
governing the rate of change of equilibrium cumulants as the system cools. 
Consider further, two distinct protocols. In the first, of type A, trajectories are very close to the Ising critical point at $r,h=0$,  corresponding to $T_c,\mu_B^c$ in the QCD phase diagram. In the Ising model, $\xieq\sim |h|^{-2/5}$ and $\theta(\tau) \sim \sgn (\ttau)$, 
for $\ttau = (\tau - \tau_c)$, with $\tau_c$ the proper time at which a trajectory crosses the crossover line at $h=0$. Near $\tau_c$, one can expand $h(\ttau) \approx \(\ttau/\tau_Q\)^a$, where $\tau_Q$ controls the rate of change of $h$ and $a$ is positive definite since $h(\tau_{c})=0$. 
Hence $\xi_{\rm eq}(\tau)\sim |\ttau/\tau_{Q}|^{-2a/5}$ and $\tqxi $ defined in \eqref{eq:quench-rates} will go to zero as $\ttau\rightarrow 0$. In contrast, $\tqth$ remains finite. Thus due to critical slowing down, for protocol A, $\tqxi \ll \teff$ very rapidly. 

\begin{figure}
\centering
\includegraphics[width=0.4\textwidth]{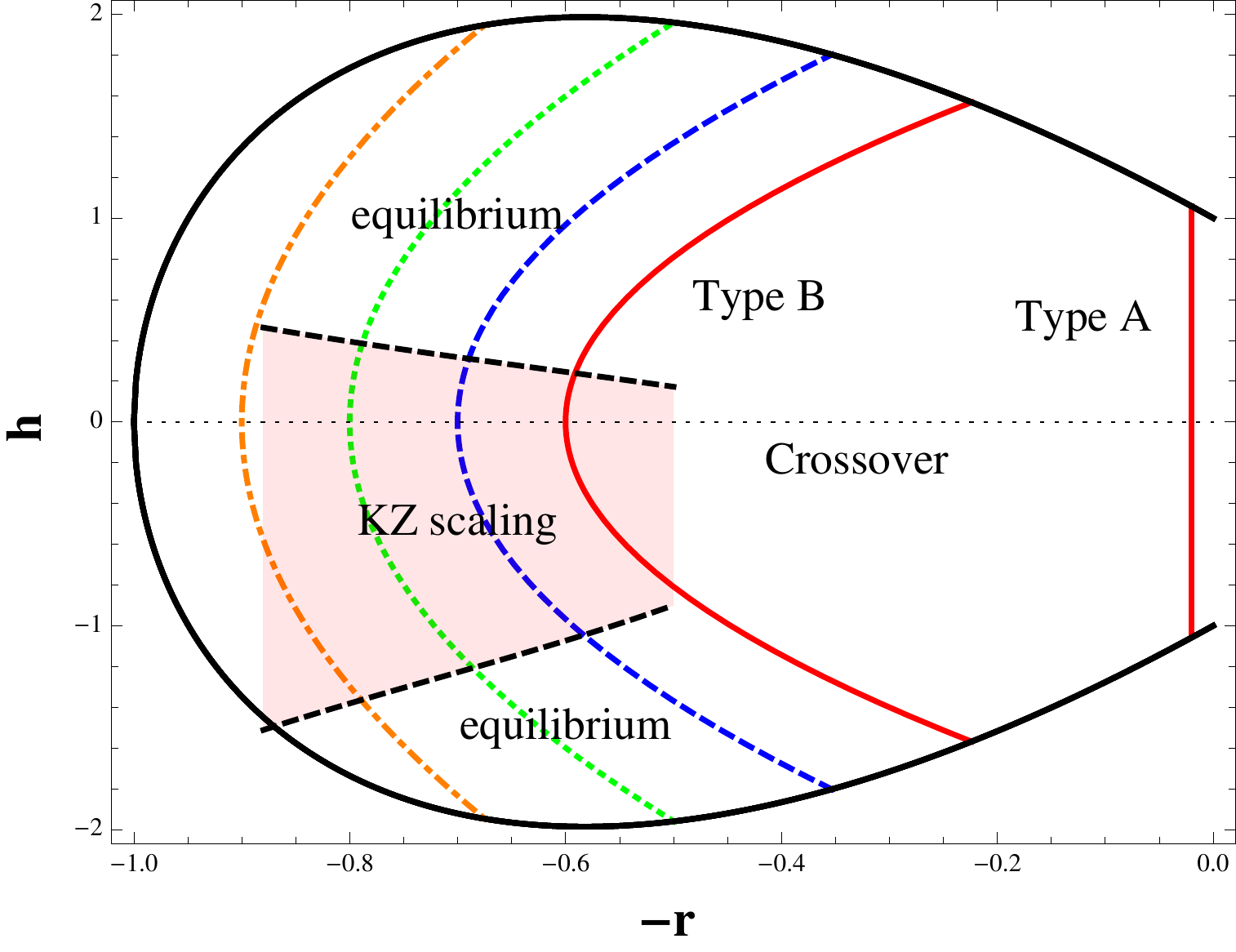}
\caption{
\label{fig:trajThetaplot}
(Color online)
Sketch of trajectories on the crossover side of the CEP. 
The solid curve delineates the boundary of the critical regime in the $r$-$h$ plane.
The rightmost trajectory represents protocol A. The other trajectories lie in protocol B.  A possible KZ scaling regime is illustrated in the shaded area.
 }
\end{figure}

We can also identify a novel protocol B for the Ising universality class. This protocol corresponds to trajectories on the crossover side ($\mu_B \leq \mu_B^c$) of the QCD phase diagram that are only weakly sensitive to critical slowing down, with $\xi_{\rm eq}(\ttau)$ reaching a maximal value at the crossover line. This implies that $\tqxi$ is large. However since $\theta$ flips sign across the crossover line, $\theta\propto \ttau$ and $\tqth$ will go to zero. Hence even though $\tqxi \gg \teff$ for protocol B trajectories, one can have $\tqth \ll \teff$. Representative trajectories in protocols A and B 
are shown in Fig.~\ref{fig:trajThetaplot}. 

The qualitative change in behavior of the quench rates relative to the relaxation rate is at the heart of the KZ dynamics. 
It allows us to define a proper time, denoted by $\tau^{*}$, at which $\tau_{\rm eff}(\tau=\tau^{*})=\tch(\tau=\tau^{*})$, 
giving rise to an emergent time scale $\tKZ$, defined through the condition,
\begin{equation}
 \label{tKZ_def}
\tKZ= 
 \t_{\rm eff}(\tau^{*})= 
 \tch(\tau^{*}) \, . 
\end{equation} 
with $\tch\equiv \min\( \tau^{\xi}_{\rm quench},  \tau^{\theta}_{\rm quench}\)$. 
One can equivalently define an emergent length scale and magnetization angle respectively to be
\begin{equation}
\label{scales}
\lKZ = \xieq  (\tau^{*}) \, ,
\qquad
 \thKZ = \theta(\tau^{*}) \, .
\end{equation}
Because critical fluctuations freeze out after $\tau^{*}$, the system retains memory of these emergent scales at later times. The equilibrium scaling of critical cumulants suggests the following ansatz:
\begin{eqnarray}
\label{eq:kappa-scaling}
\kappa_{n}(\tau;\G)
\sim \lKZ^{-\frac{1}{2}+\frac{5}{2}(n-1)}\, {\bar f}^{I}_{n}(t;\theta_{\rm KZ})\, ,
\end{eqnarray}
with $t = \ttau / \tKZ$ and $I$ labels different protocol classes. 
While $\tKZ$, $\lKZ$ and $\thKZ$ depend non-universally on $\G$, the functions ${\bar f}^{I}_{n}$ are universal for all the trajectories characterizing a given protocol. A possible regime of protocol B where such scaling may hold is sketched in Fig.~\ref{fig:trajThetaplot}.

\begin{figure}
\center
\subfigure[]
{
\includegraphics[width=0.4\textwidth]{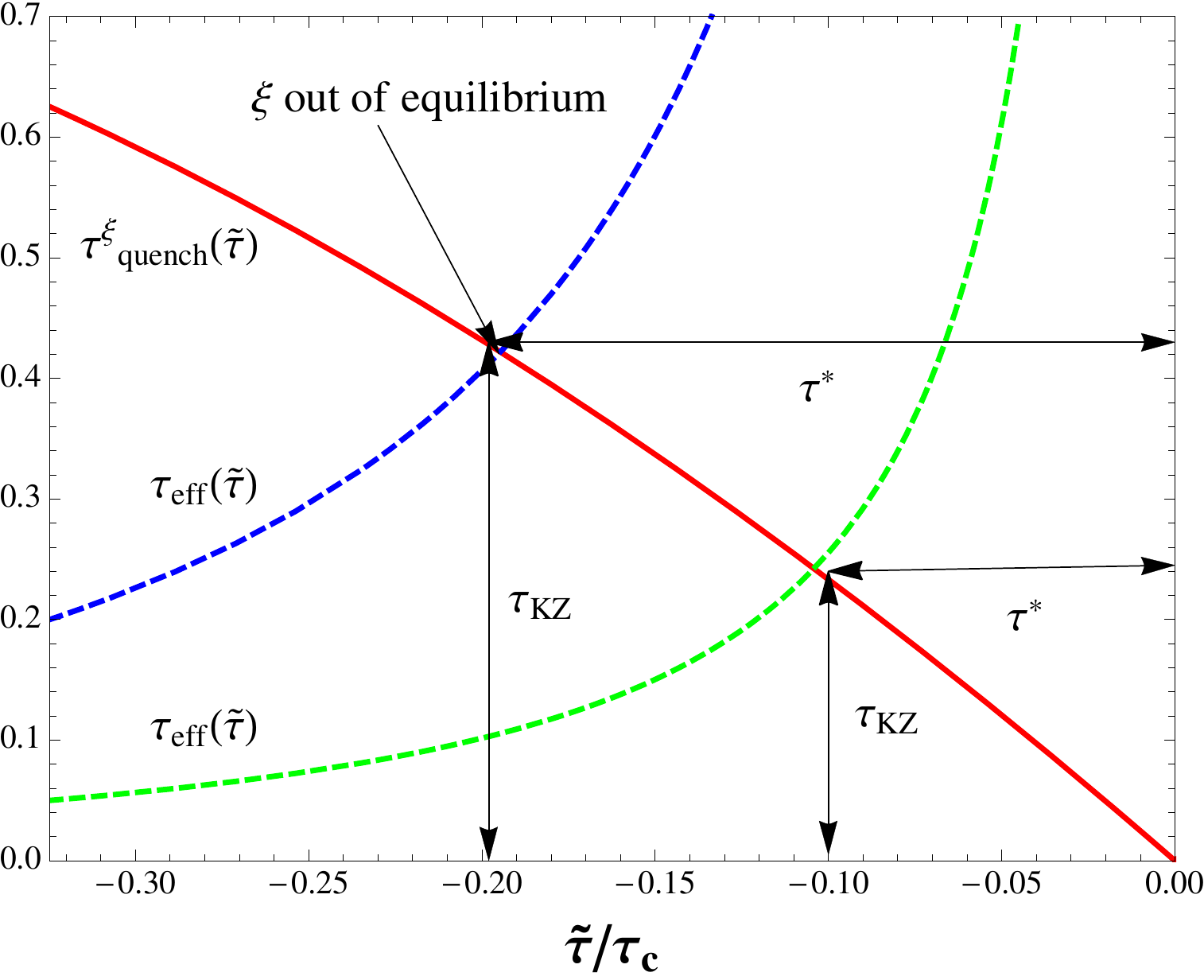}
\label{fig:xiquench}
}
\subfigure[]
{
\includegraphics[width=0.4\textwidth]{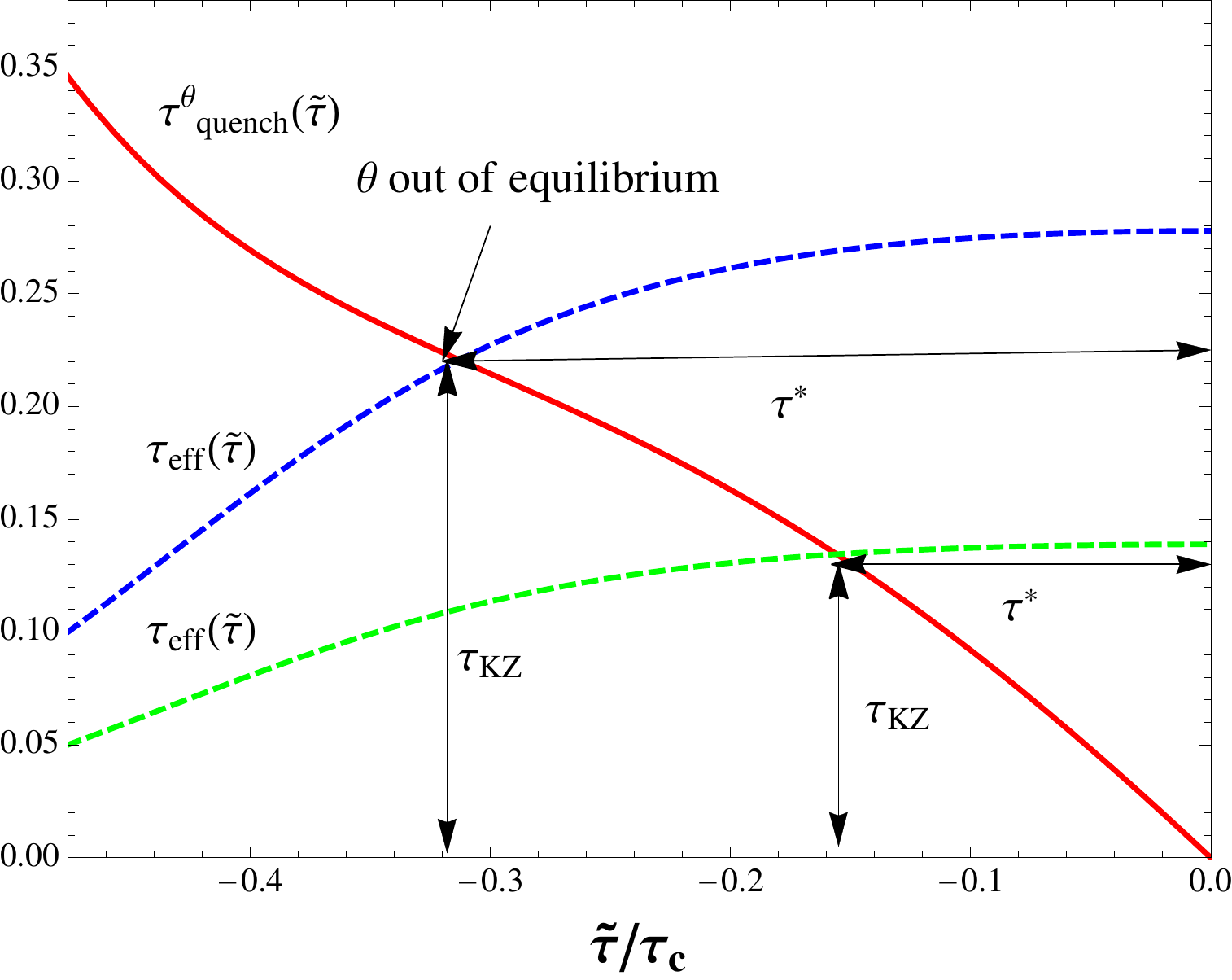}
\label{fig:thetaquench}
}
\caption{
\label{fig:quenchtime}
(Color online)
(a): the evolution of $\tau^{\xi}_{\rm quench}$ and $\tau_{\rm eff}$ with two different choices of $\trel$ along representative trajectories of protocol A. 
(b): the evolution of $\tau^{\theta}_{\rm quench}$ and $\tau_{\rm eff}(\ttau)$, likewise, along representative trajectories of protocol B. 
 }
\end{figure}
In Fig.~\ref{fig:xiquench}, we plot the temporal evolution of $\tqxi$ for a characteristic quench scale $\tau_Q$ we will specify later and compare it to $\teff = \trel (\xi/\xmin)^3$, where $\trel$ and $\xmin$ are the relaxation time of the critical mode and equilibrium correlation length respectively at the boundary of the critical regime. 
For the two different $\trel$ s along a trajectory in  protocol A,
we obtain distinct values of $\tKZ$ when $\teff$ crosses $\tqxi$; one can also straightforwardly extract $\lKZ$. For protocols B, Fig.~\ref{fig:thetaquench} shows that one similarly obtains a $\tKZ$ that corresponds to an novel KZ magnetization angle $\thKZ$. Note that $\trel, \xmin$ are non-universal parameters that are part of $\Gamma$ and $\tKZ,\lKZ,\thetaKZ$ depend on $\G$.

In Appendix A, we present analytical arguments that justify the scaling form in Eq.~(\ref{eq:kappa-scaling}) for both protocols. 
However, one can use the closed form expressions~\cite{Mukherjee:2015swa} for $\kappa_{n}$ 
to check numerically the existence and domain of validity of the scaling. Towards this end, 
we will adopt a widely used but non-universal map~\cite{Berdnikov:1999ph,Nonaka:2004pg}  between the Ising and QCD parameters, wherein $(T- T_c)/\Delta T=h$ and $(\mu_{B}-\mu^{c}_B)/\Delta \mu_{B}= -r$, with $\Delta T, \Delta \mu$ denoting the width of the critical regime in the QCD phase diagram. 
(The normalization of $ r, h$ are fixed by the conditions $\xi(r=1, h=0)=\xi(r=0,h=1) = \xi_{\min}$.)
For the fireball in heavy ion collisions, we will use $T= T_c [\tau/\tau_c]^{-3 c_s^2}$, with the temperature evolution of the three dimensional isentropic expansion \footnote{It is easy to check that $a=1$ and  $\tau_Q = \tau_c (\Delta T/T_c)/(3 c_s^2)$.} determined by the speed of sound $c_s$. 

\begin{figure}
\centering
\subfigure[]
{
\includegraphics[width=0.4\textwidth]{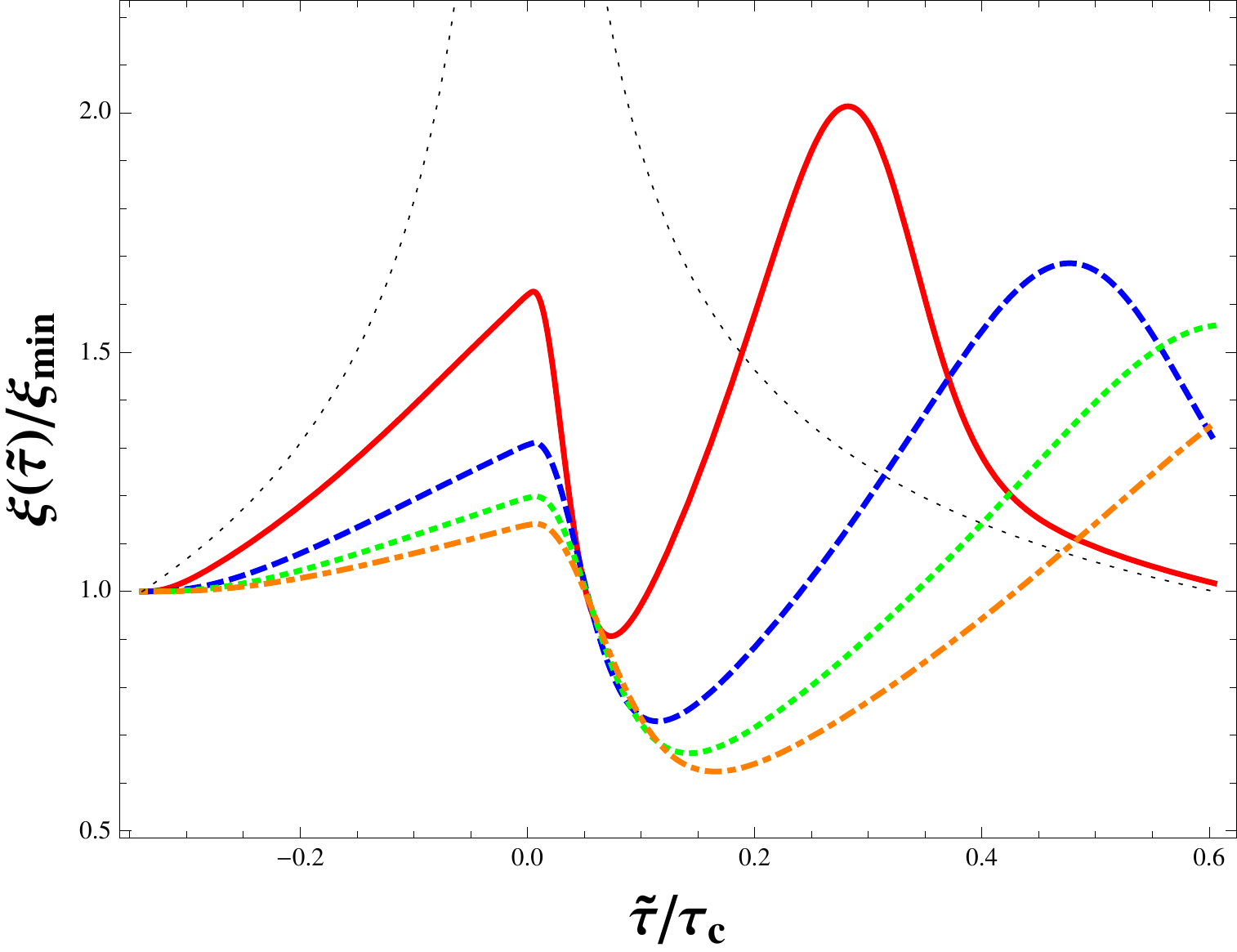}
\label{fig:xiplotA}
}
\subfigure[]
{
\includegraphics[width=.4\textwidth]{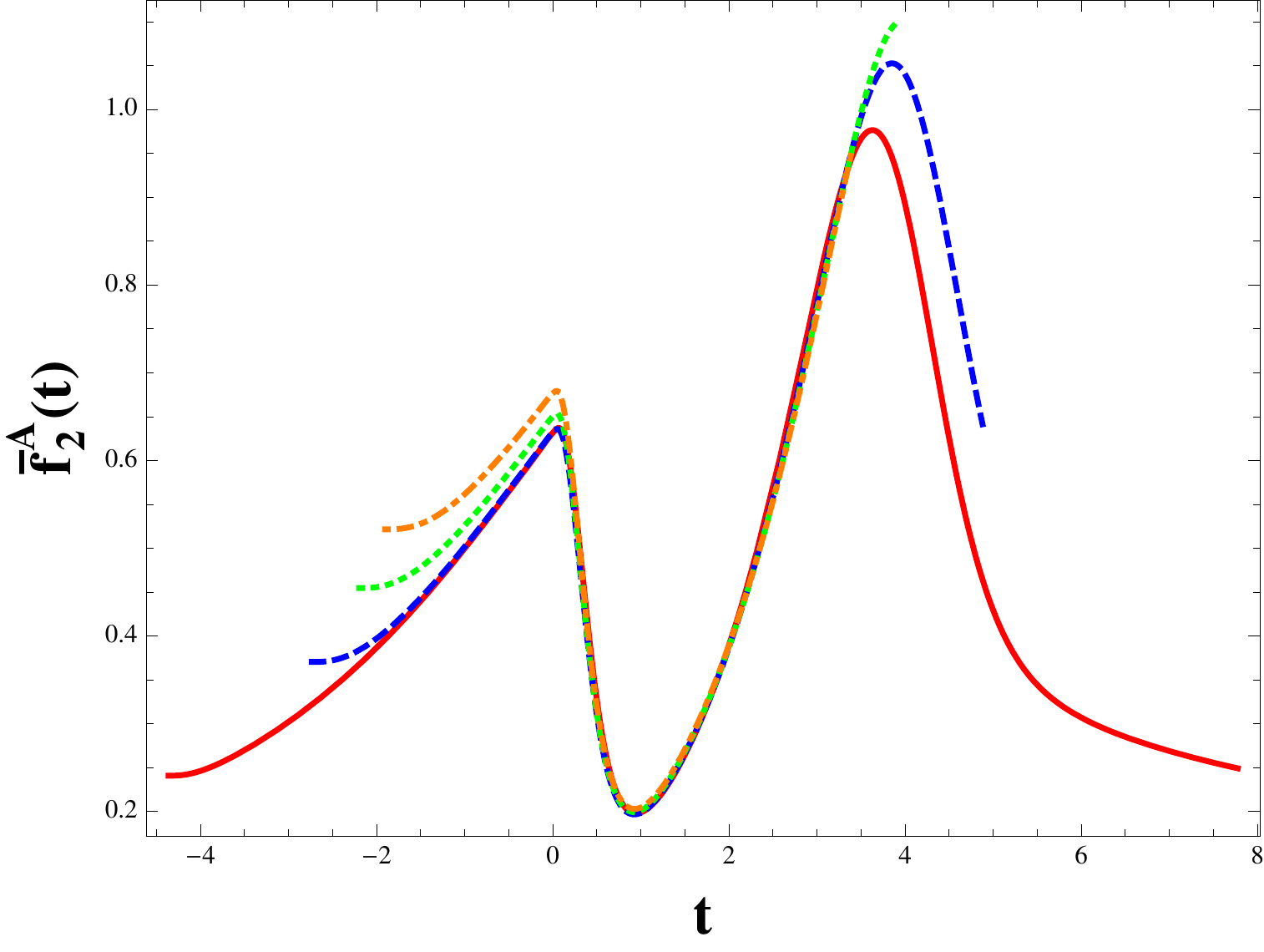}
\label{fig:f2plotA}
}
\caption{
\label{fig:secondA}
(a)
: the evolution of the non-equilibrium effective correlation length  $\xi(\tilde{\tau})/\xi_{\rm min}$ for protocol A.
The corresponding equilibrium value is plotted in dotted curve. 
(b)
: the rescaled function ${\bar f}^{A}_{2}(t)$ vs the rescaled time $t=\ttau/\tKZ$. 
Results with $\tau_{\rm rel}/\tau_{c}=0.02,0.06,0.1,0.14$ are shown in red, dashed blue, dotted green and dot-dashed orange curves respectively.
 }
\end{figure}

In Fig.~\ref{fig:xiplotA}, we plot the non-equilibrium correlation length $\xi$ over $\xmin$ for different choices of $\trel$ in protocol A. 
The trajectory for each such choice is clearly non-universal and varies significantly with $\trel$. 
Now using Eq.~(\ref{eq:kappa-scaling}) and constructing $\tKZ$ as specified, we plot the  function $\bar{f}^{A}_2$ as a function of $t$. 
As anticipated by our scaling ansatz, it scales beautifully; the different curves in Fig.~\ref{fig:f2plotA}, obtained by solving the cumulant equation in Ref.~\cite{Mukherjee:2015swa}  
for $\kappa_2$, 
collapse onto a single nearly universal curve.
Equally impressive scaling is seen for the magnetization ($\kappa_1$), 
skewness $\kappa_{3}$ and kurtosis $\kappa_{4}$.
The equivalent protocol A plots for these are respectively shown in 
Figs.~\ref{fig:firstA}, \ref{fig:thirdA} and \ref{fig:fourthA}
of Appendix~\ref{sec:test}.

\begin{figure}
\centering
\subfigure[]
{
\includegraphics[width=0.4\textwidth]{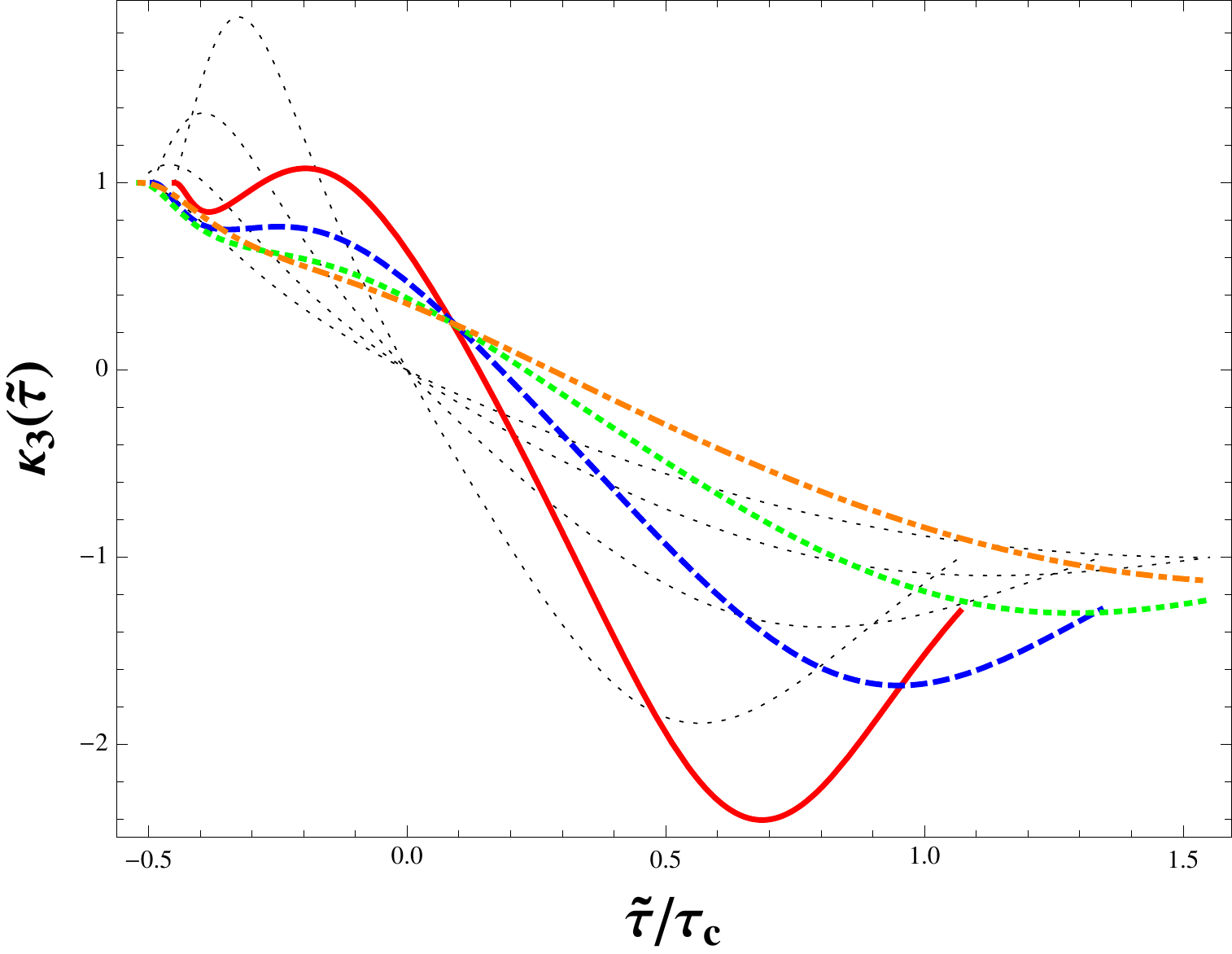}
\label{fig:kappa3plotB}
}
\subfigure[]
{
\includegraphics[width=.4\textwidth]{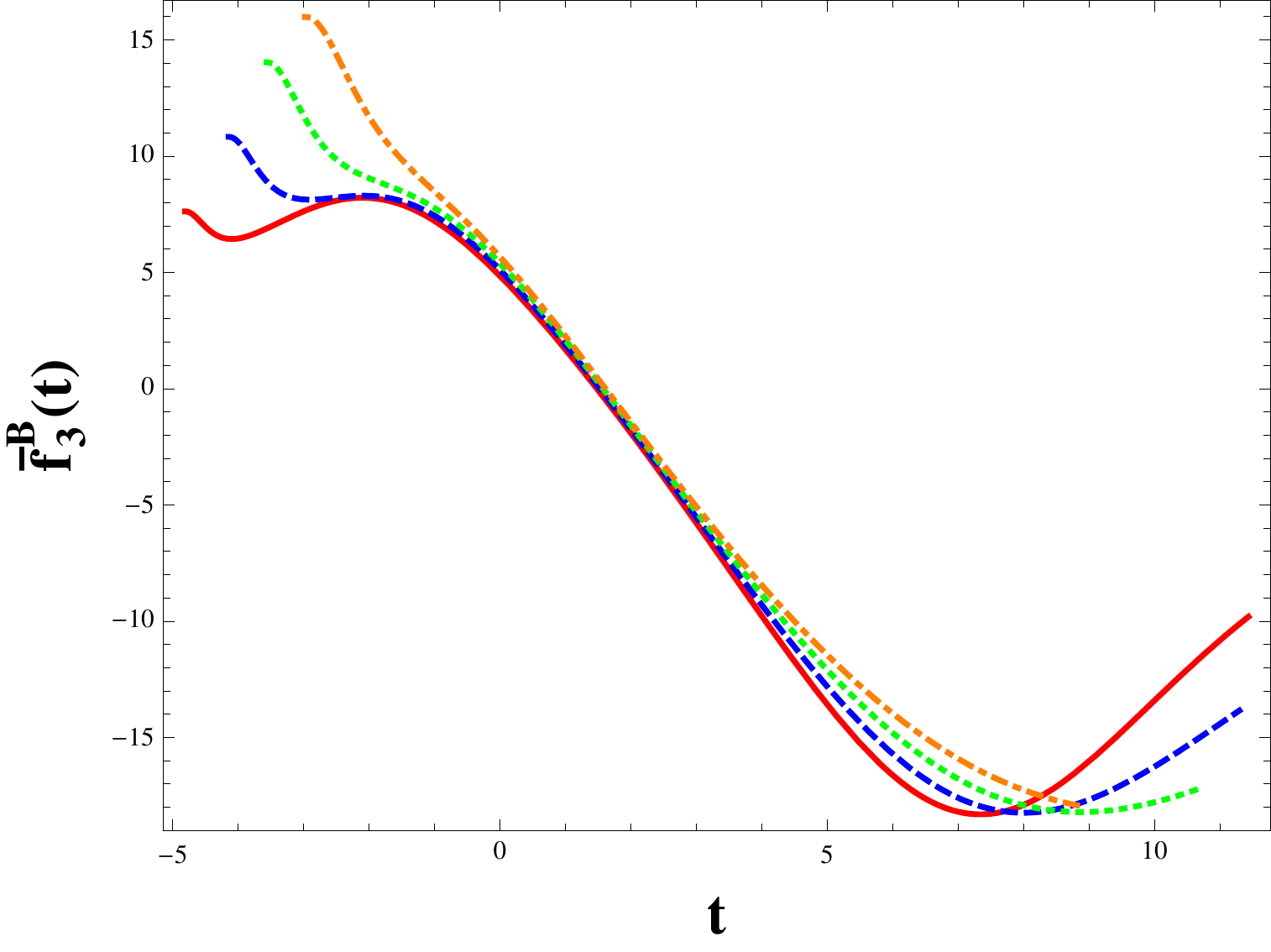}
\label{fig:f3plotB}
}
\caption{
\label{fig:thirdB}
(a)
: Nonequilibrium evolution of $\kappa_{3}(\tilde{\tau})$ (normalized by its initial equilibrium value) for representative trajectories in protocol B. 
The corresponding equilibrium values are plotted in dotted curves. 
(b)
: the rescaled function ${\bar f}^{B}_{3}(t)$ versus the rescaled time $t=\tau/\tKZ$.
The red, blue dashed, green dotted and orange dot-dashed curves correspond to those shown in Fig.~\ref{fig:trajThetaplot}. 
 }
\end{figure}

\begin{figure}
\centering
\subfigure[]
{
\includegraphics[width=0.4\textwidth]{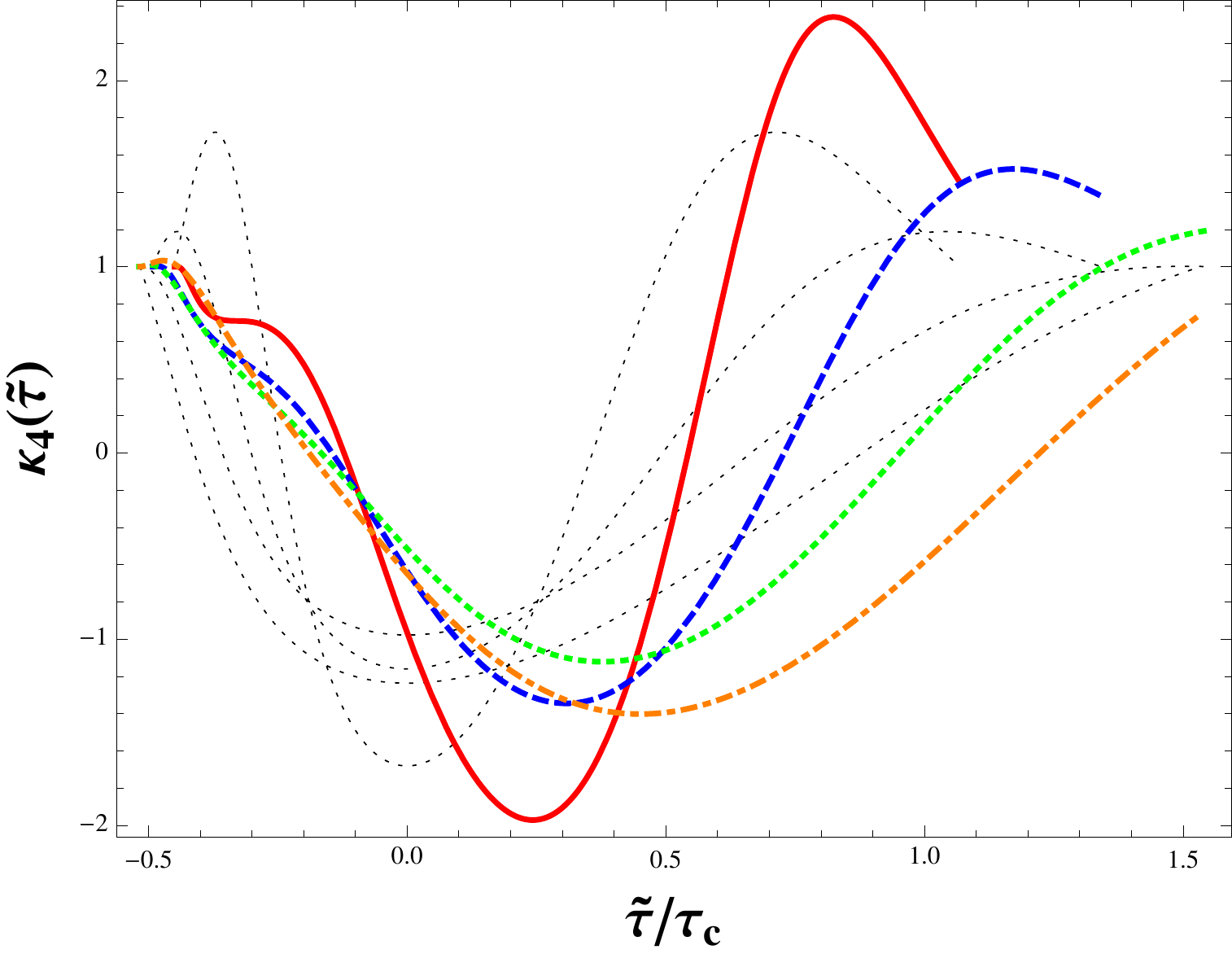}
\label{fig:kappa4plotB}
}
\subfigure[]
{
\includegraphics[width=.4\textwidth]{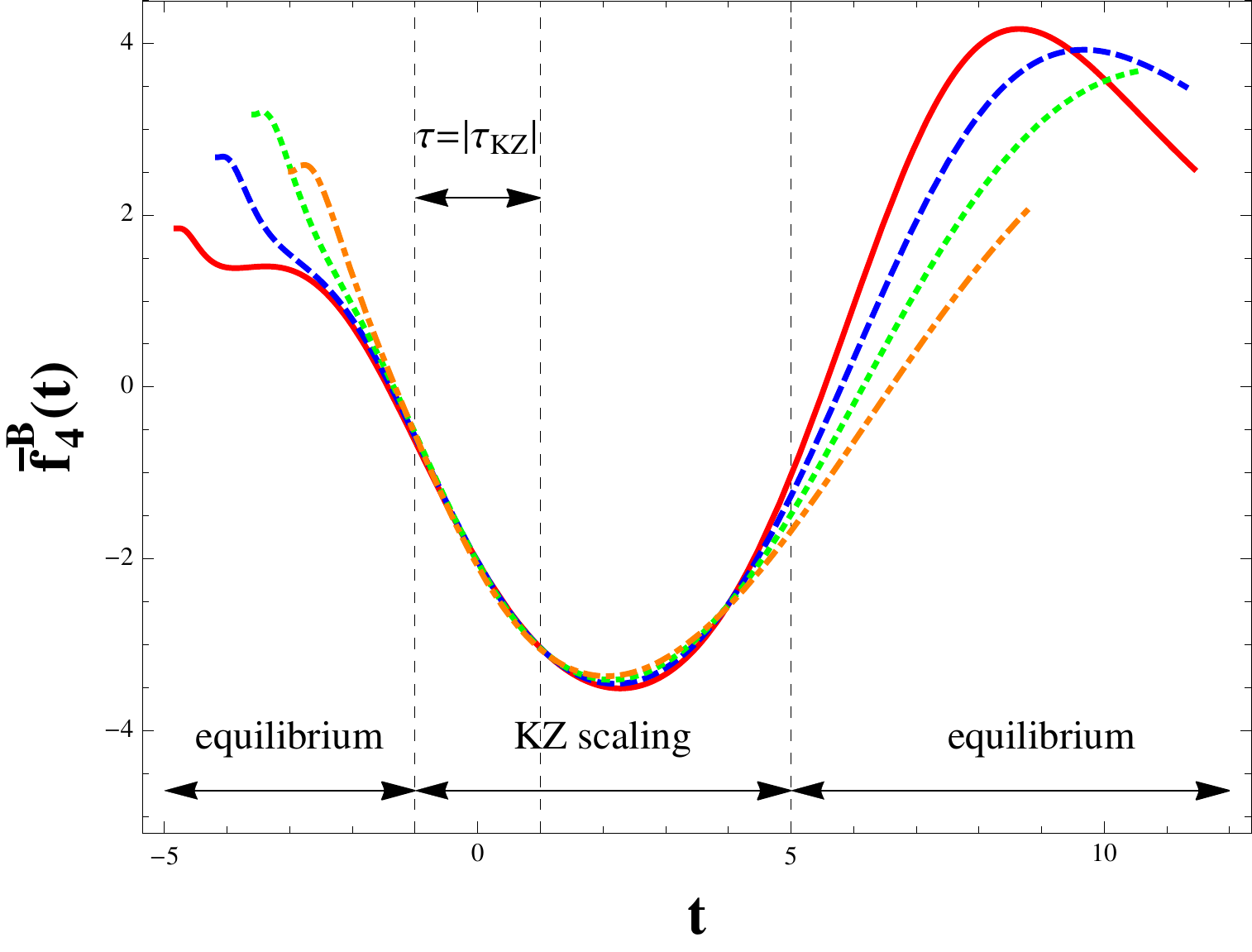}
\label{fig:f4plotB}
}
\caption{
\label{fig:fourthB}
Evolution of $\kappa_4$ and ${\bar f}^{B}_4$ for protocol B trajectories along the lines described in the caption for Fig.~\ref{fig:thirdB}.
 }
\end{figure}

Turning now to protocol B, we will examine the behavior of the four trajectories shown in Fig.~\ref{fig:trajThetaplot}. 
We tune $\trel$ in such a way that $\thKZ$ is identical ($\thKZ=-0.1$) for the evolution along each trajectory.
In Figs.~\ref{fig:kappa3plotB} and \ref{fig:kappa4plotB}, we show the corresponding cumulants $\kappa_3$ and $\kappa_4$ obtained from solving cumulant equation in Ref.~\cite{Mukherjee:2015swa}. 
Following the same procedure as for protocol A, we plot the functions ${\bar f}^{B}_{3}$, ${\bar f}^{B}_{4}$ as a function of t in in Figs.~\ref{fig:f3plotB} and \ref{fig:f4plotB}. 
Very good scaling is observed in both cases, confirming the validity of our hypothesis. One naively expects the non-equilibrium scaling hypothesis to only apply in the regime $|\ttau|<\tau_{\rm KZ}$ (or $|t|<1$). This is because the critical cumulants will approach their corresponding equilibrium values outside the KZ regime. 
Our numerical results for both protocols demonstrate that the KZ scaling solution persists for much longer, suggesting that the KZ scaling functions are attractor solutions. 
For a discussion of the latter, see Ref.~\cite{Berges:2009jz}. 

We will now consider what these findings imply for the BES search for the CEP in the QCD phase diagram. 
An immediate consequence is that if BES trajectories are sensitive to the critical point in some window of $\sqrt{s}$ (center of mass), the centrality (degree of overlap), and rapidity in the collisions, cumulants of hadron multiplicity distributions sensitive to the critical modes~\cite{Hatta:2003wn,Athanasiou:2010kw,Asakawa:2015ybt} should be expressible in the scaling form suggested by Eq.~(\ref{eq:kappa-scaling}). 
In particular, if the KZ scaling regime is probed by the freeze-out curve of hadrons emitted at proper time $\ttau_{f}$ from the BES fireballs,  the critical cumulants, after rescaling with the appropriate powers of $l_{\rm KZ}$, will only depend on $\thKZ$ and $t_{f}\equiv \ttau_{f}/\tau_{KZ}$ for trajectories in the same protocol. How can the search for KZ scaling be achieved in practice? 

The steps we propose are as follows: 
\begin{enumerate}
\item
Hydrodynamic modeling of the evolution of bulk properties in the BES, with appropriate choices of initial conditions, should be able to quantitatively reproduce spectra of various hadron species~\cite{Nahrgang:2016ayr,Karpenko:2015xea}. 
Thus details of a given trajectory can be determined by such modeling; for each such trajectory one can use Eq.~(\ref{eq:quench-rates}) to extract the quench times for the variables $\xi$ and $\theta$ controlling the critical dynamics. 
\item 
Determine $\tKZ$ by solving Eq.~(\ref{tKZ_def}) and compute likewise $\lKZ$ and $\thKZ$ from Eq.~(\ref{scales}) for this trajectory.  
Read $\ttau_{f}$ from the position of the freeze-out curve in the hydro simulation. 
We note that $\tKZ$, $\lKZ$, $\thKZ$ still depend on a subset of non-universal inputs from critical properties of QCD matter that we denote as $\G_{\rm crit}$: these are the aforementioned mapping between between $r,h$ and $T,\mu_{B}$, the location of the critical point $\mu^{c}_{B}, T_{c}$, the width of the critical regime $\Delta \mu_{B}, \Delta T$ and $\tau_{\rm rel}$.
\item
Compute rescaled cumulant data of observables sensitive to critical dynamics as ${\bar f}^{\rm data}_{n}\equiv \kappa^{\rm data}_{n}/l^{-\frac{1}{2}+\frac{5}{2}(n-1)}_{\rm KZ}$. One can than establish a mapping between $\kappa^{\rm data}_{n}$ to a point in 
$\({\bar f}^{\rm data}_{n}, t_{f}, \thKZ\)$ space. 
We note from the previous step that this mapping depends on $\G_{\rm crit}$.
\item 
Repeat the above steps for windows in $\sqrt{s}$, centrality and rapidity that are sensitive to critical dynamics.  Data on the corresponding cumulants mapped to $\({\bar f}^{\rm data}_{n}, t_{f}, \thKZ\)$ space should collapse onto a single surface by suitably adjusting $\G_{\rm crit}$. 
This surface will be described by the scaling functions ${\bar f}_{n}(t, \thKZ)$.

\item
In parallel to the previous steps, compute the universal scaling functions by solving the cumulant equations along \textit{one} representative trajectory of each protocol. 
Compare the theoretically computed ${\bar f}_{n}(t_{f},\theta_{\rm KZ})$ with rescaled data to further confirm the scaling hypothesis.  
\end{enumerate}
If such theory-data comparisons are successful, they would provide unambiguous evidence for the existence of the QCD CEP\footnote{While presence of KZ scaling in data would strongly indicate existence of a CEP, in contrast, absence of scaling would not necessarily indicate the absence of a CEP. Fireball trajectories may miss the critical region entirely, may not traverse it for very long, or KZ freeze-out may be destroyed if significant further evolution of the fireball occurs.}. 
The analysis sketched above should also allow us to extract $\G_{\rm crit}$, which encodes important properties of QCD matter near the CEP. 
The procedure outlined, with examples including mock BES data, will be pursued in future work. 
It can also be explored in models that explicitly couple critical and bulk dynamics--along the lines of previous work~\cite{Herold:2016uvv}.

There are a number of features of our results that are of broader interest. 
The non-equilibrium scaling of non-Gaussian cumulants has received little attention in the literature on the KZ dynamics. 
A noteworthy exception is an approach based on the reparametrization invariance~\cite{Nikoghosyan:2013fqa} of the stochastic master equations representing the mathematical content of different dynamical universality classes~\cite{RevModPhys.49.435}. This approach has much in common with our analytical discussion of the structure of cumulants in the Appendix.~A and may provide deeper insight into the wider than expected KZ scaling regime observed. 
Another novel observation is that the quench rate $\tqth$ for a class of Ising  trajectories (our protocol B) can be rapid enough that magnetization angle can freeze-out; this result is of particular importance for higher cumulants that are sensitive to the sign of this angle. Finally, it will be interesting to explore the merits of alternative mechanisms for the non-equilibrium critical dynamics of gauge theories~\cite{Hindmarsh:2000kd,Rajantie:2001ps} relative to the Kibble-Zurek framework explored here.

\section*{Acknowledgment} 
The authors would like to thank J.~Berges, P.~Braun-Munzinger, U.~Heinz, M.~Hindmarsh, B.-L. Hu, A.~Kovner, J.~Pawlowski and K.~Rajagopal for useful discussions. RV thanks the Institut f\"{u}r Theoretische Physik, Heidelberg for their kind hospitality and the Excellence Initiative of Heidelberg University for support.
This material is partially based upon work supported by the U.S. Department of Energy, Office of Science, Office of Nuclear Physics, under contract No. DE- SC0012704, and within the framework of the Beam Energy Scan Theory (BEST) Topical Collaboration.

\begin{appendix}

\section{Analytic insights into scaling hypothesis}
\label{sec:analytic}

In this appendix, we will show analytically 
the existence of scaling solution \eqref{eq:kappa-scaling} for off-equilibrium evolution of critical cumulants near the crossover line.
Our starting point is the evolution equations derived in Ref.~\cite{Mukherjee:2015swa} for critical cumulants. 
This set of evolution equations is equivalent to the Fokker-Planck equation which describes the evolution of the probability distribution of critical modes. 
For an alternative derivation of the Kibble-Zurek mechanism based on the reparametrization invariance of the Fokker-Planck equation, see also  Ref.~\cite{Nikoghosyan:2013fqa}. 

We now begin our discussion by rewriting evolution equation for cumulants (Eq.~2.20,  in Ref.~\cite{Mukherjee:2015swa}) in a more general form:
\begin{eqnarray}
\label{eq:kappa_evo}
\pd_{\tau}\kappa_{n}(\tau)
&=& -
\frac{n}{\tau_{\eff}}\, 
\tF_{n}\[\theta, \e, b;\kappa_{1},\ldots,\kappa_{n} \]\, ,
\end{eqnarray}
where we have introduced two parameters:
\begin{equation}
\label{b-def}
\epsilon\equiv \sqrt{\frac{\xi^{3}_{\rm eq}}{V}}\, , 
\qquad
b= \sqrt{\frac{\xi^{2}_{\equ}T}{V}}\, .
\end{equation}
The functions $\tF_{n}$ for $n=1,2,3,4$ are given in Ref.~\cite{Mukherjee:2015swa}--for the reader's convenience, we collect them at the end of this Appendix in Eqs.~\eqref{tF}. 

Motivated by the non-equilibrium scaling hypothesis~\eqref{eq:kappa-scaling},
we consider the following ansatz for the evolution of cumulants:
\begin{eqnarray}
\label{mod-kappa-scaling}
\kappa_{n}(\tau)= C^{-2+n}_{0}\, \frac{T^{\frac{1}{2}n}_{c}}{ V^{n-1}_{c}}\, l^{-\frac{1}{2}+\frac{5}{2}(n-1)}_{\rm KZ}f_{n}(\tau;\G)\, 
\end{eqnarray}
Here
$T_{c},V_{c}$ denote the temperature and volume when the system passes the crossover line. 
$C_{0}$ is a non-universal normalization constant, 
see also Eqs.~\eqref{tF}. 

We now substitute \eqref{mod-kappa-scaling} into Eq.~\eqref{eq:kappa_evo}. 
In this work, 
we will replace $V, T$ in~\eqref{b-def} with $V_{c},T_{c}$ respectively. 
This amounts to replacing $\e$ with $\e_{c}=\sqrt{\xi^{3}_{\rm eq}/V_{c}}$ and $b$ with $b_{c}=\sqrt{\xi^{2}_{\rm eq}T_{c}/V_{c}}$. 
Such a simplification is justified as long as the quench time of the ratio of the temperature/volume is much longer than $\tau^{\xi}_{\rm quench}$ or $\tau^{\theta}_{\rm quench}$. 
As a result of these substitutions, Eq.~(\ref{eq:kappa_evo}) can be written as
\begin{eqnarray}
\label{f-evo}
\tau_{\rm KZ}\pd_{\tau}f_{n}(\tau;\G)
&=& - \frac{n}{\tilde{\tau}_{\rm eff}}\, 
G_{n}\[\tilde{\xi}_{\rm eq},\theta; f_{1},\ldots, f_{n}\]\, , 
\end{eqnarray}
where we introduced the scaled ratios
\begin{equation}
\label{A-B-def}
\tilde{\tau}_{\rm eff}(\tau)\equiv \frac{\tau_{\rm eff}(\tau)}{\tau_{KZ}} \, , \qquad
\tilde{\xi}_{\rm eq}(\tau)\equiv \frac{\xi_{\equ}(\tau)}{l_{KZ}}\, . 
\end{equation}
The {\it functional form} of $G_{n}\[\tilde{\xi}_{\rm eq},\theta; f_{1},\ldots, f_{n}\]$ is universal; for convenience, we only list the somewhat cumbersome expressions at the end of this Appendix--in Eqs.~(\ref{G-fun}). By observation, $G_{n}$ only depends explicitly on $\tilde{\xi}_{\rm eq},\theta$ and $f_{1},\ldots, f_{n}$.
Therefore if $\tilde{\tau}_{\rm eff}, \tilde{\xi}_{\rm eq}, \theta$ only depend on the rescaled time $t=\tau/\tau_{\rm KZ}$ and $\theta_{\rm KZ}$, 
the scaling form $\bar{f}(t;\theta_{\rm KZ})$ will solve Eq.~(\ref{f-evo}). 

We will now check explicitly for trajectories belong to protocol A or protocol B, whether the evolution of $\tilde{\tau}_{\rm eff}, \tilde{\xi}_{\rm eq}, \theta$ in the vicinity of the crossover line indeed depends only on $t, \theta_{\rm KZ}$. 
If so, this would confirm the existence of universal scaling solutions. 

Let us first consider protocol A.
Near $T_{c}$, one could use expansion  $\xieq\sim |h|^{-2/5}$ and therefore we have $\tau^{\xi}_{\rm quench}\approx \frac{5}{2}|\tau|$ and
the condition in Eq.~(\ref{tKZ_def}) to determine $\tau_{\rm KZ}$ becomes 
\begin{equation}
  \label{tKZ_2}
 \tau_{\rel} \Big |\frac{\tau^{*}}{\tau_{Q}}\Big|^{-\frac{6}{5}}
 = \frac{5}{2}|\tau^{*}|\, ,
\end{equation}
where we have used $\tau_{\rm eff}=\tau_{\rm rel}\(\frac{\xi}{\xi_{\rm min}}\)^{3}$. 
Likewise, $\lKZ$ can be determined from Eq.~(\ref{scales}) and one can check that
\bes
\label{t-A}
\begin{equation}
\label{AB-t-A}
\tilde{\xi}_{\rm eq}(t) \approx |\frac{5}{2}t|^{-2/5}\, , 
\qquad
\tilde{\tau}_{\rm eff} \approx \tilde{\xi}^{3}_{\rm eq}\, ,
\end{equation}
for evolution near $T_{c}$.
Turning now to $\theta(\tau)$, 
we found from $\theta(\tau) \sim \sgn (\ttau)$ and the definition of $\theta_{KZ}$ in Eq.~(\ref{scales}) that
\begin{equation}
\label{theta-t-A}
\theta(\tau) \approx \theta_{KZ}\, \sgn(t)\, . 
\end{equation}
\ees
This concludes our proof for protocol A that Eq.~(\ref{f-evo}) has a scaling solution of the form $\bar{f}^{A}_{n}(t;\theta_{KZ})$ near $T_c$.

We next consider protocol B. Since for protocol B, $\xi_{\rm eq}$ reaches its maximum when crossing the crossover line, 
we have $\xi_{\rm eq}\sim l_{\rm KZ}$ and $\tau_{\rm eff}\sim \tau_{\rm KZ}$ for evolution in the vicinity of the crossover line. 
Thus $\tilde{\xi}_{\rm eq}\approx 1, \tilde{\tau}_{\rm eff}\approx 1$. 
On the other hand, since $\theta \propto \tilde{\tau}\equiv \tau-\tau_{c}$,
we will have from Eq.~(\ref{scales}),
\begin{equation}
\theta(\tau) \approx \theta_{\rm KZ} t\,  , 
\end{equation}
for protocol B.  We therefore conclude that Eq.~(\ref{f-evo}) has a scaling solution of the form $\bar{f}^{B}_{n}(t;\theta_{KZ})$.
Since $ \tau_{\rm eff}(t), \xi_{\rm eff}(t),\theta(t)$ take different forms for protocol A and B, $\bar{f}^{A}_{n}$ and $\bar{f}^{B}_{n}$ correspond to distinct 
universal scaling functions.

\begin{figure}
\centering
\subfigure[]
{
\includegraphics[width=0.4\textwidth]{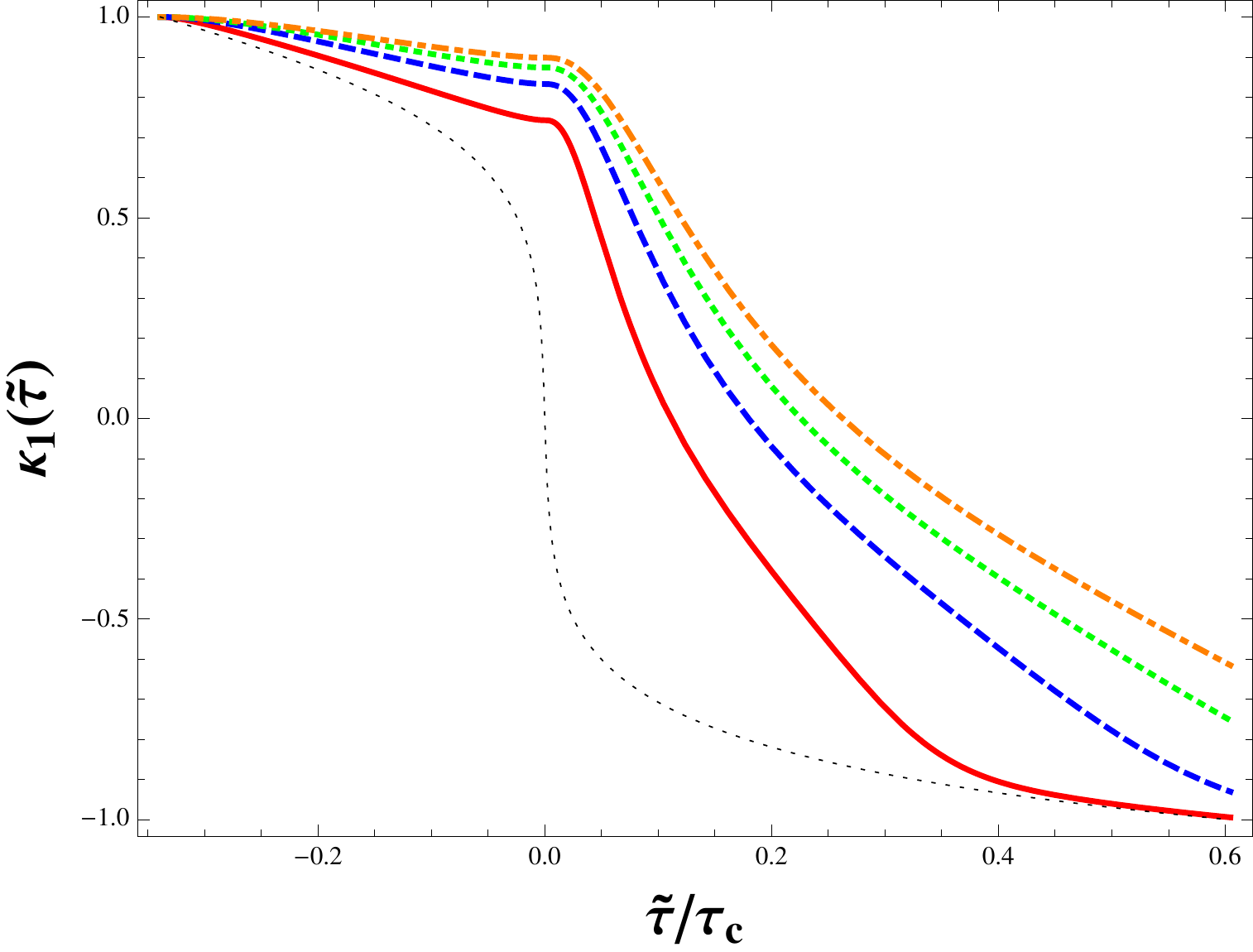}
\label{fig:kapp1plotA}
}
\subfigure[]
{
\includegraphics[width=.4\textwidth]{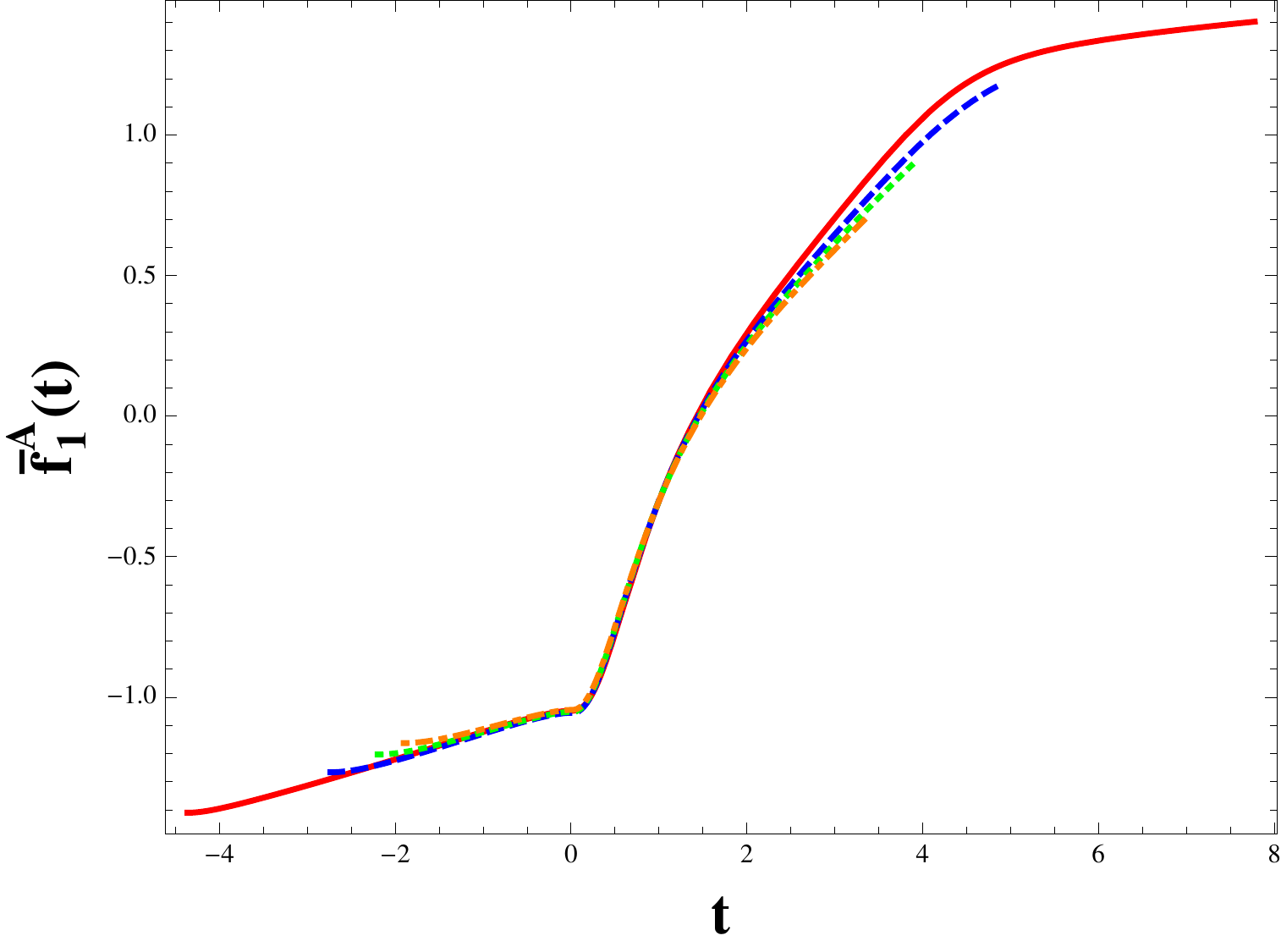}
\label{fig:f1plotA}
}
\caption{
\label{fig:firstA}
The evolution of the non-equilibrium ``magnetization''  $\kappa_{1}(\tilde{\tau})$ (a) and $\bar{f}^{A}_{1}$ (b)
for a representative protocol A trajectory along the lines described in the caption for Fig.~\ref{fig:secondA}.
 }
\end{figure}

\begin{figure}
\centering
\subfigure[]
{
\includegraphics[width=0.4\textwidth]{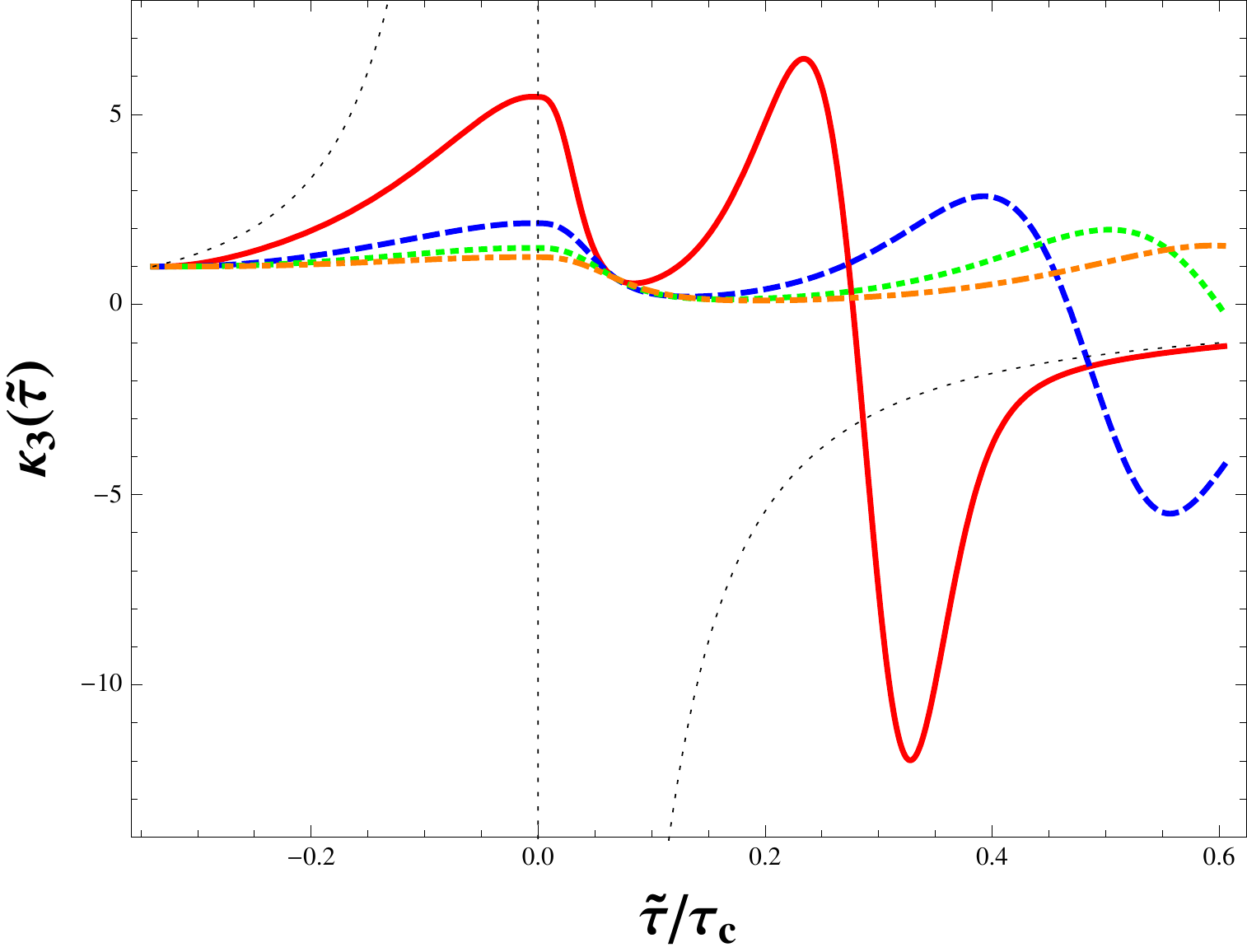}
\label{fig:kappa3plotA}
}
\subfigure[]
{
\includegraphics[width=.4\textwidth]{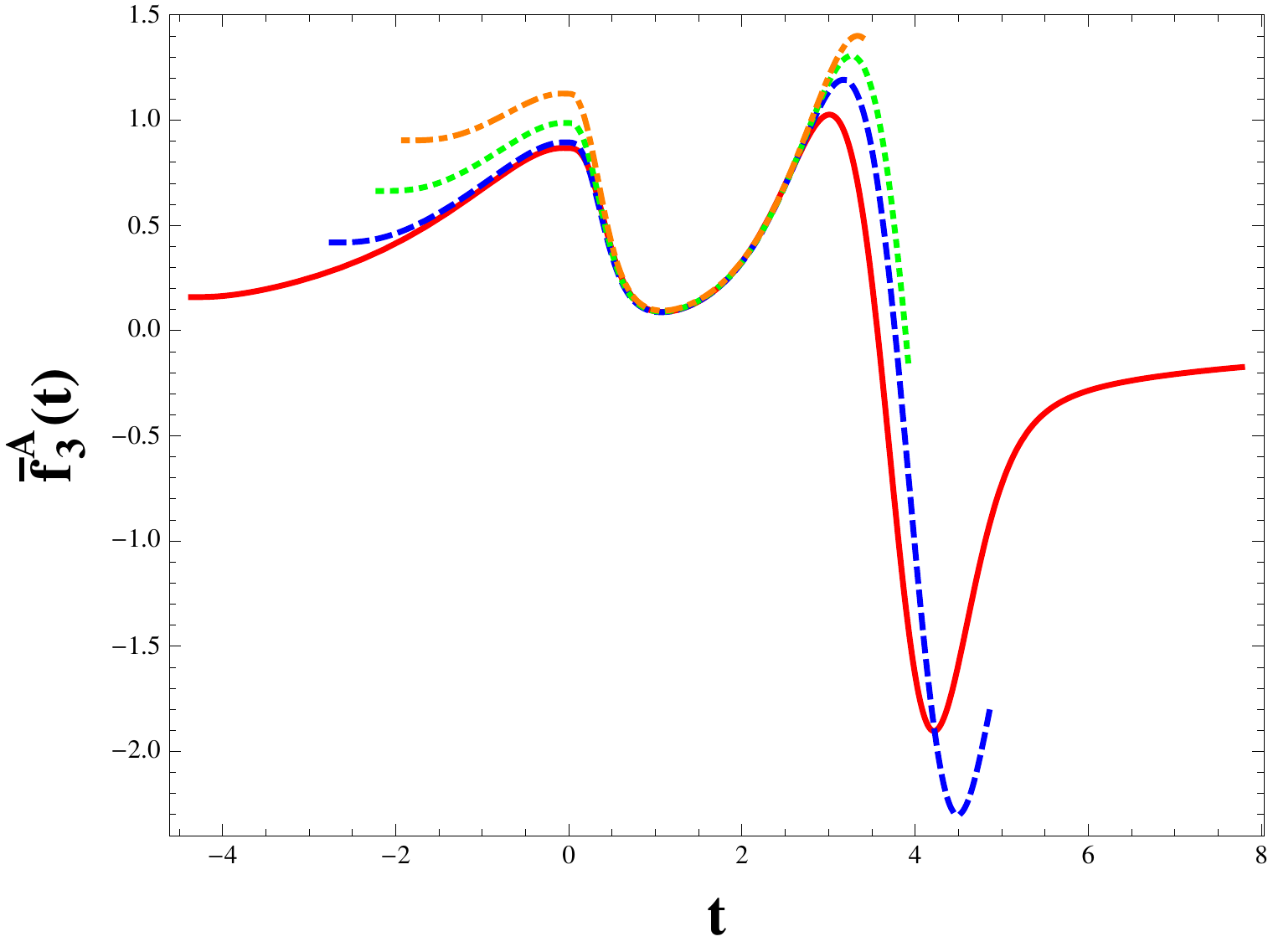}
\label{fig:f3A}
}
\caption{
\label{fig:thirdA}
The evolution of  $\kappa_{3}(\tilde{\tau})$ (a) and $\bar{f}^{A}_{3}$ (b)
for a representative protocol A trajectory along the lines described in the caption for Fig.~\ref{fig:secondA}.
 }
\end{figure}

\begin{figure}
\centering
\subfigure[]
{
\includegraphics[width=0.4\textwidth]{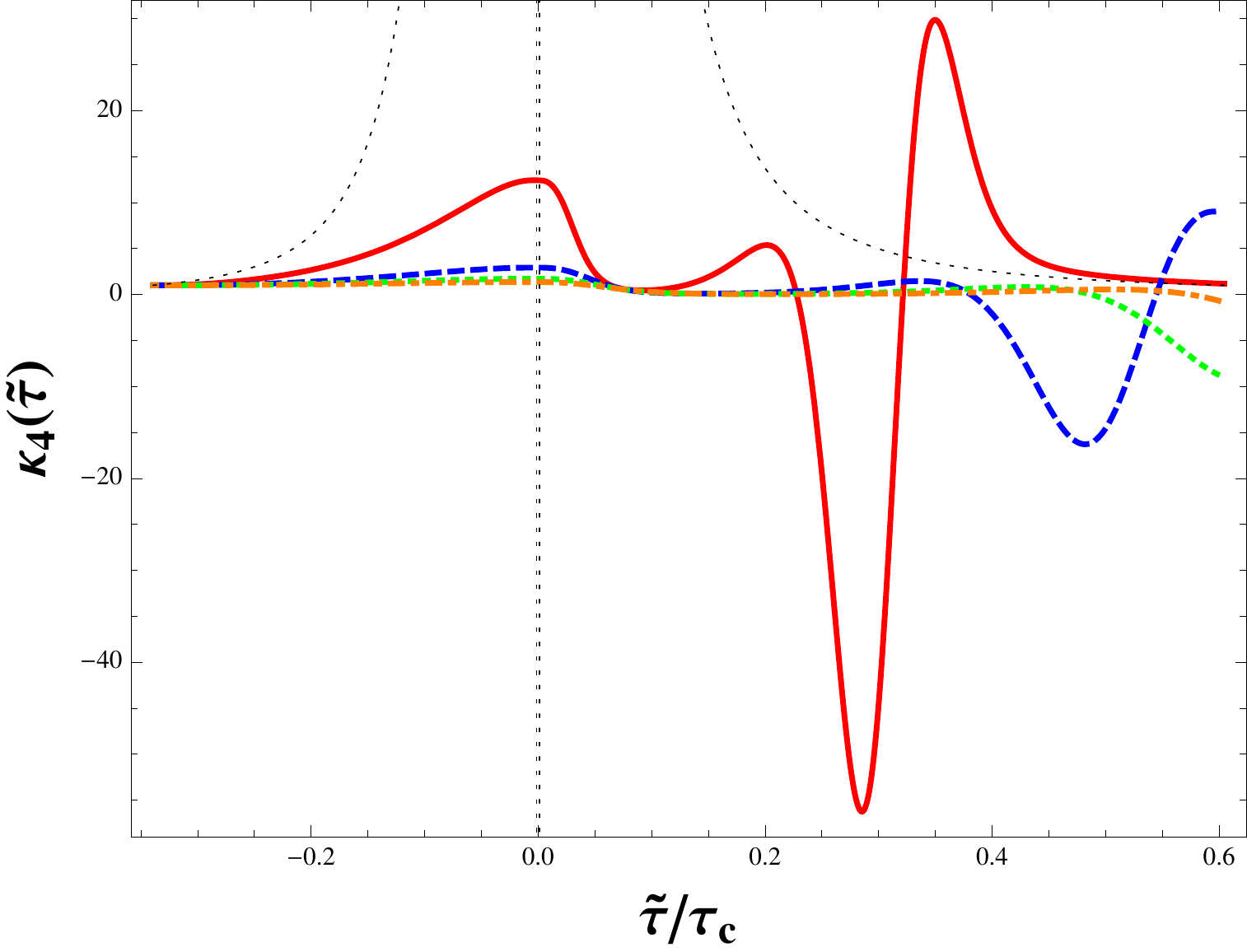}
\label{fig:kappa4plotA}
}
\subfigure[]
{
\includegraphics[width=.4\textwidth]{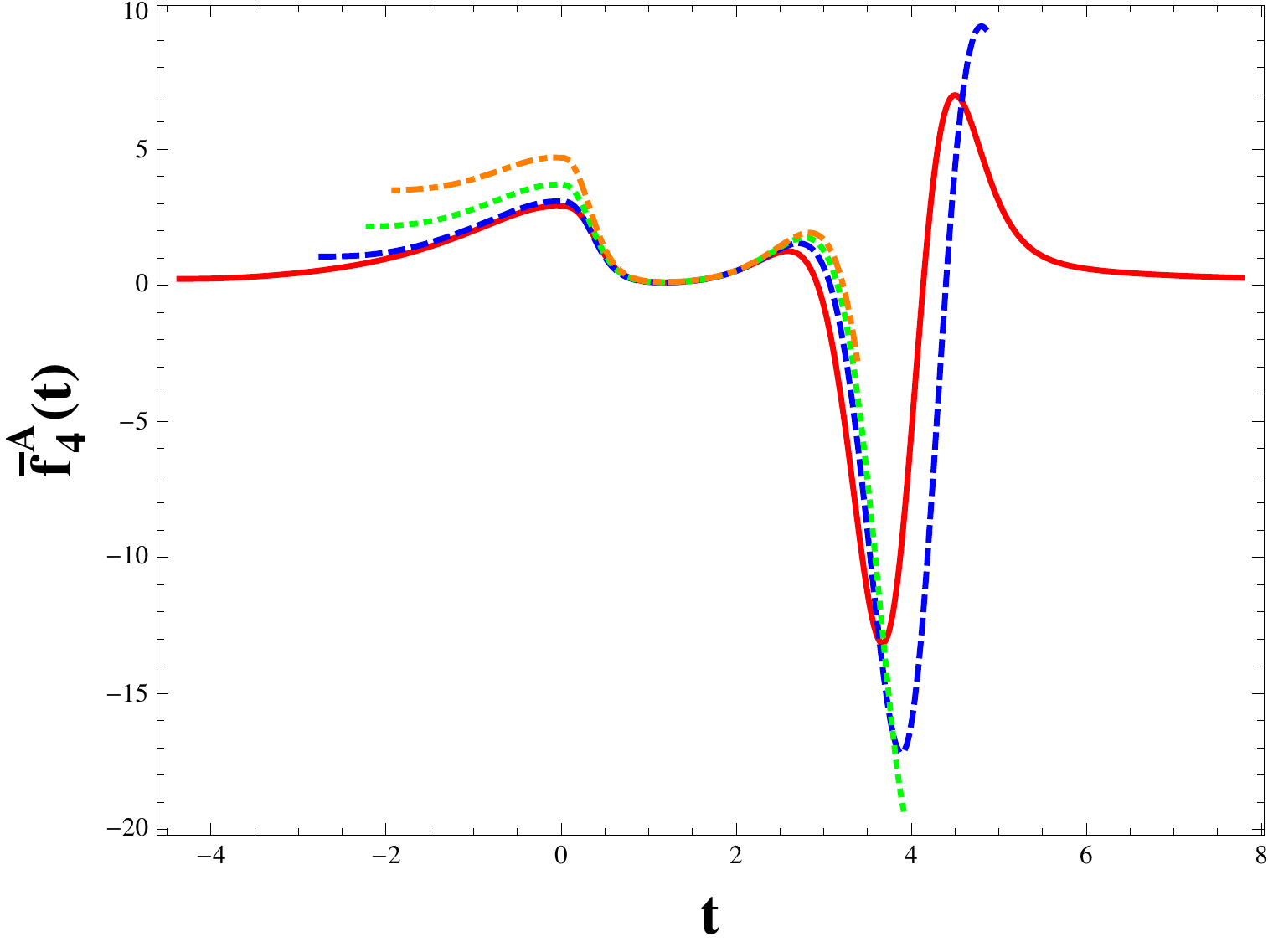}
\label{fig:f4plotA}
}
\caption{
\label{fig:fourthA}
The evolution of  $\kappa_{4}(\tilde{\tau})$ (a) and $\bar{f}^{A}_{4}$ (b)
for a representative protocol A trajectory along the lines described in the caption for Fig.~\ref{fig:secondA}.
}
\end{figure}
We conclude this section by collecting explicit expressions for the right hand side of Eq.~(\ref{eq:kappa_evo}) and Eq.~(\ref{f-evo}) for $n=1,2,3,4$. 
Following Ref.~\cite{Mukherjee:2015swa}, $\tF_{n}$ in \eqref{eq:kappa_evo} reads
\begin{eqnarray}
\label{tF}
\tF_{1}
\Big[\theta,\e,b;\kappa_{1}\Big] &=& 
 \d \tM\Big[1+ \tl_{3}(\theta) (\d \tM)
 \nonumber\\
&+& \tl_{4}(\theta)(\d \tM)^2\Big]
 \, ,
\nonumber\\
\tF_{2}\Big[\theta,\e,b;\kappa_{1},\kappa_{2}\Big]&=&
\(b^2\) \, \[ \(\frac{\k_{2}}{b^2}\)F_{2}\(\delta \tM;\theta\)  -1\]
\nonumber\\
\tF_{3}\Big[\theta,\e,b;\kappa_{1},\kappa_{2},\kappa_{3}\Big]
&=&-\(\e \,b^{3}\)
\Big[ \( \frac{\k_{3}}{\e \,b^3}\)F_{2}\(\delta \tM;\theta\) 
\nonumber\\
&+&
 \(\frac{\k_{2}}{b^2}\)^2F_3\(\delta \tM;\theta\)\Big]
\, , 
\nonumber \\
\tF_{4}\Big[\theta,\e,b;\kappa_{1},\kappa_{2},\kappa_{3},\kappa_{4}\Big]
&=&
\Bigg\{\, 
\(\frac{\k_{4}}{\e^2\,b^4}\) F_{2}\(\delta \tM;\theta\)
\nonumber\\
&+&3\(\frac{\k_{2}}{b^2}\)\(\frac{\k_{3}}{\e \,b^3}\) F_{3}\(\delta \tM;\theta\)
\nonumber\\
&+&
6\(\frac{\k_{2}}{b^2}\)^3 \tl_{4}(\theta)
\Bigg\}\, . 
\end{eqnarray}
As in Ref.~\cite{Mukherjee:2015swa}, 
$\delta\tM, F_{2,3}$ are given by
\begin{eqnarray}
\label{eq:F_exp}
\delta \tM\[\theta,\e,b;\kappa_{1}\,\]
&=&
\Big [\,\(\frac{\e}{b}\)\kappa_{1}-C^{-1}_{0}\, \tilde{\sigma}(\theta)\,\Big]\, , 
\nonumber\\
F_{2}(\d\tM)
&=&1+ 2 \tl_{3}(\theta)(\d \tM)+3\tl_{4}(\theta)(\d \tM)^2\, ,
\nonumber\\
F_{3}(\d\tM) 
&=&2 \[\tl_{3}(\theta)+3\tl_{4}(\theta) (\d \tM)\]
\, ,
\end{eqnarray}
and $\tsigma_{0}(\theta), \tl_{3}(\theta),\tl_{4}(\theta)$ are determined from a linear parametrization model of the Ising equation of state~\cite{PhysRevLett.23.1098,ZinnJustin:1999bf}:
\begin{eqnarray}
\label{tilde-theta}
\tsigma_{0}(\theta)&=&  \,\frac{5^{1/4}\theta}{\(3+2\theta^{2}\)^{1/4}}\, , 
\\
\tl_{3}(\theta)&=&\frac{1}{5^{1/4}}\,\frac{2\theta (9+\theta^{2})}{\(3-\theta^{2}\)\(3+2\theta^{2}\)^{3/4}}\, , 
\\
\tl_{4}(\theta)&=&\frac{1}{\sqrt{5}}\,\frac{2\(27+45\theta^{2}-31\theta^{4}-\theta^{6}\)}{\(3-\theta^{2}\)^{3}\(3+2\theta^{2}\)^{1/2}}
\end{eqnarray}
Here the dimensionless quantity $C_{0}$ is non-universal. 

We next consider $\tG_{n}$ which appears in \eqref{f-evo}. 
By straightforward calculation, 
we have:
\begin{eqnarray}
\tG_{1}\[\tilde{\xi}_{\rm eq},\theta;f_{1}\]&=& G_{1}\, \[ f_{1} ;\tilde{\xi}_{\rm eq}, \theta\] \,
G_{0}\[ f_{1} ;\tilde{\xi}_{\rm eq}, \theta\]\, 
\nonumber\\
\tG_{2}\[\tilde{\xi}_{\rm eq},\theta;f_{1},f_{2}\]&=& \,\tilde{\xi}_{\rm eq} \,
\Big\{
\, G_{2}\[ f_{1}; \tilde{\xi}_{\rm eq}, \theta \]f_{2}-\tilde{\xi}_{\rm eq}
\Big\}\, , 
\nonumber\\
\tG_{3}\[\tilde{\xi}_{\rm eq},\theta;f_{1},f_{2},f_{3}\]
&=&
\tilde{\xi}_{\rm eq}\, \Big\{
 G_{2}\[ f_{1}; \tilde{\xi}_{\rm eq}, \theta \]f_{3}  \nonumber \\
 &+&
 2\,  G_{3}\[ f_{1}; \tilde{\xi}_{\rm eq}, \theta\] f^{2}_{2}
\Big\}\, , 
\nonumber\\
\tG_{4}\[\tilde{\xi}_{\rm eq},\theta;f_{1},f_{2},f_{3}\]
&=&  \tilde{\xi}_{\rm eq}\, \Big\{ G_{2}\[ f_{1}; \tilde{\xi}_{\rm eq}, \theta \]\, f_{4}
\nonumber \\
 &+& 6\, G_{3}\[ f_{1}; \tilde{\xi}_{\rm eq}, \theta \]\, f_{2}\,f_{3} 
 \nonumber\\
&+& 6\tl_{4}(\theta) f^{3}_{2} \Big\}\, . 
\end{eqnarray}
where
\begin{eqnarray}
\label{G-fun}
G_{0}\[ f_{1} ;\tilde{\xi}_{\rm eq}, \theta\]
&=&\[\, f_{1}-\tilde{\xi}_{\rm eq}^{-1/2}\tilde{\sigma}(\theta)\, \]  \, , 
\nonumber \\
G_{1}\[ f_{1}; \tilde{\xi}_{\rm eq}, \theta\]
&=&\[ \tilde{\xi}_{\rm eq}^{-1}+ \tilde{\xi}_{\rm eq}^{-1/2}\,\tl_{3}(\theta)\, G_{0}
+ \tl_{4}(\theta)\,\(G_{0}\)^{2}\]\, ,
\nonumber\\
G_{2}\[ f_{1}; \tilde{\xi}_{\rm eq}, \theta\]&=&\Big[ \tilde{\xi}_{\rm eq}^{-1}+ 2\,\tilde{\xi}_{\rm eq}^{-1/2}\tl_{3}(\theta)\, G_{0} \,  +3\, \tl_{4}(\theta)\(G_{0}\)^{2}\Big]\, ,
\nonumber\\
G_{3}\[ f_{1}; \tilde{\xi}_{\rm eq}, \theta\]
&=&
 \Big[ \tilde{\xi}_{\rm eq}^{-1/2}\tl_{3}(\theta)+ 3\tl_{4}(\theta)\, G_{0} \Big]\, . 
\end{eqnarray}

\section{More detailed numerical results for trajectories A and B}
\label{sec:test}

\begin{figure}
\centering
\subfigure[]
{
\includegraphics[width=0.4\textwidth]{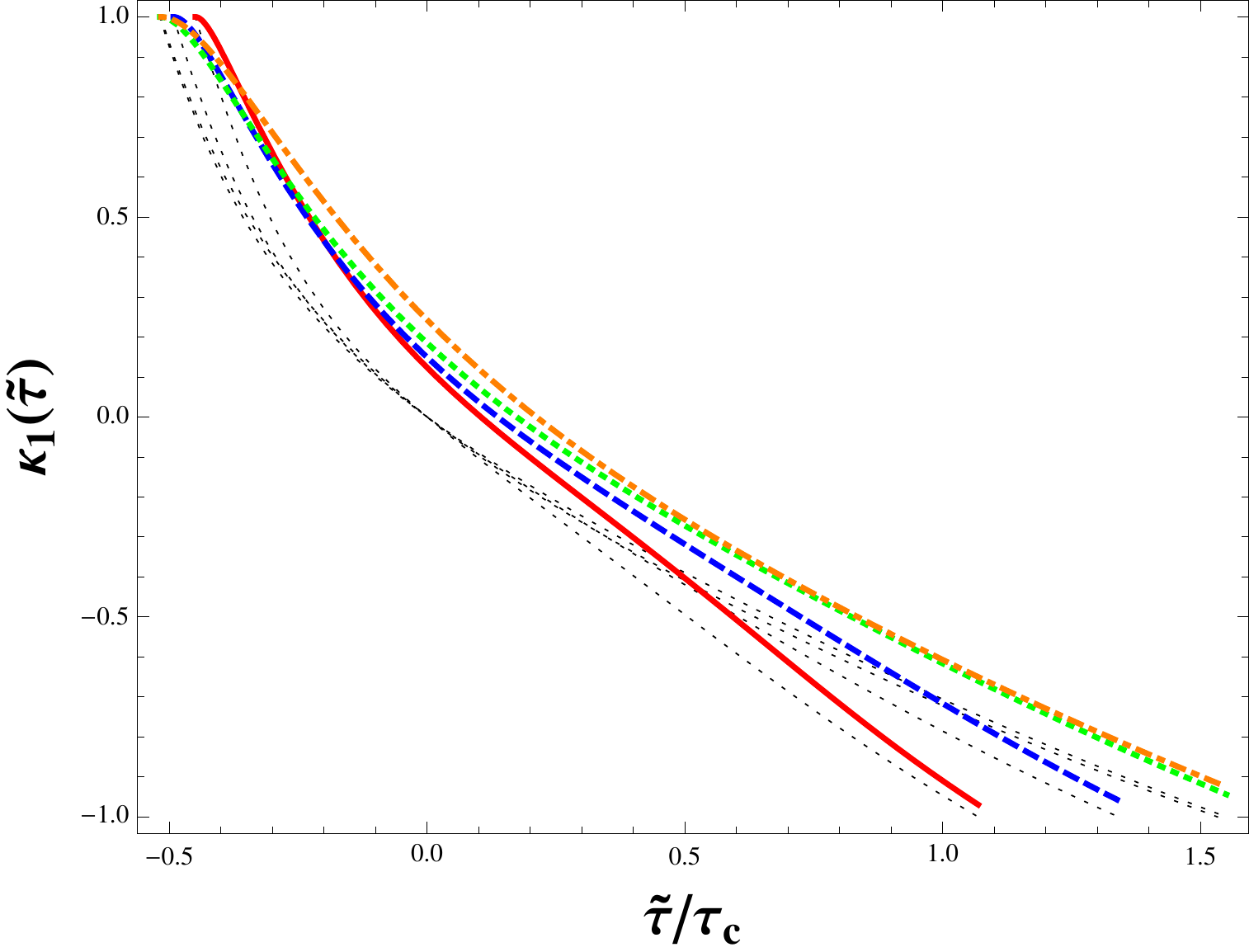}
\label{fig:kappa1plotB}
}
\subfigure[]
{
\includegraphics[width=.4\textwidth]{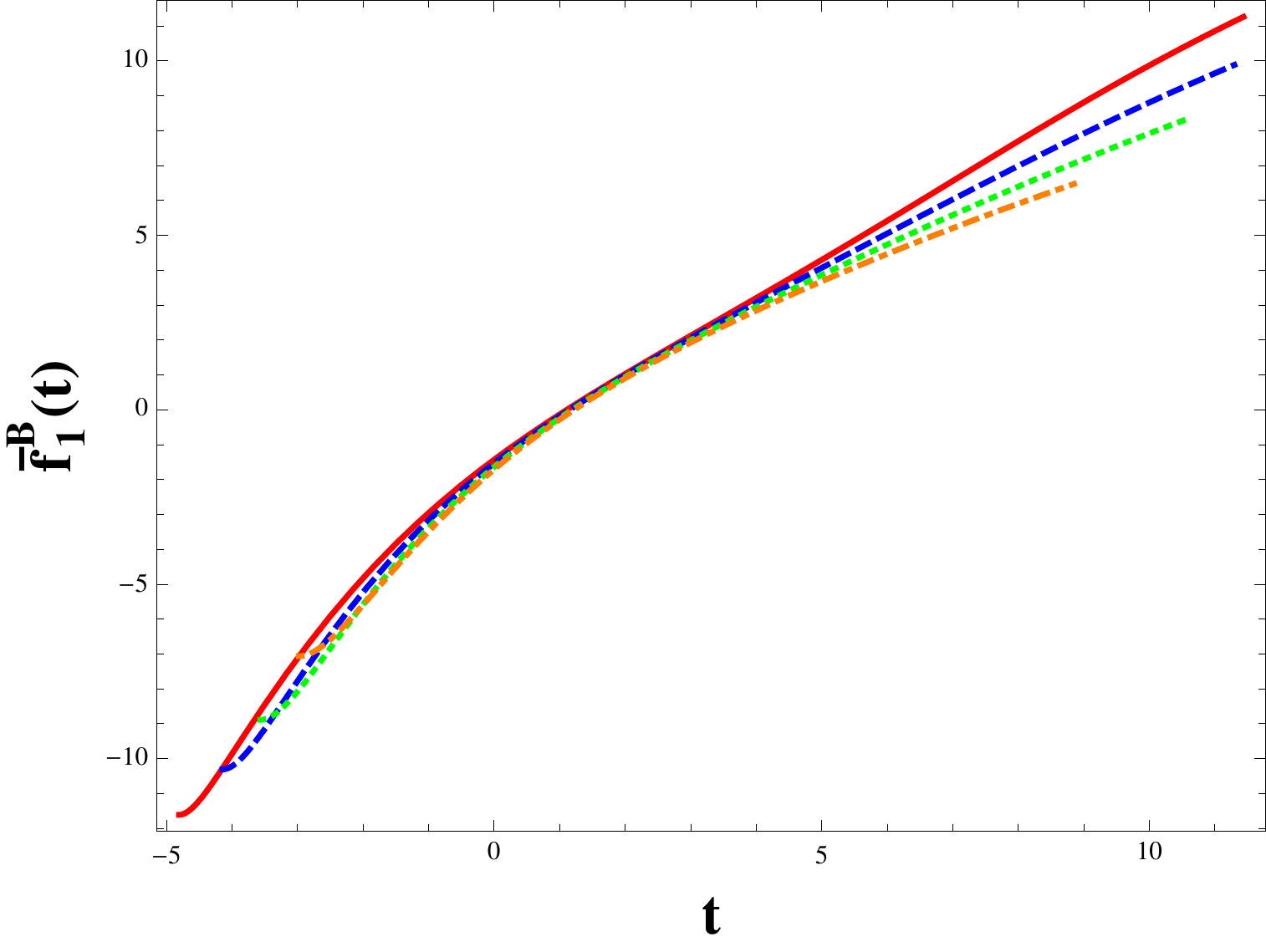}
\label{fig:f1plotB}
}
\caption{
\label{fig:firstB}
Evolution of $\kappa_{1}$ (a) and ${\bar f}^{B}_1$ (b) for protocol B trajectories along the lines described in the caption for Fig.~\ref{fig:thirdB}.
 }
\end{figure}

\begin{figure}
\centering
\subfigure[]
{
\includegraphics[width=0.4\textwidth]{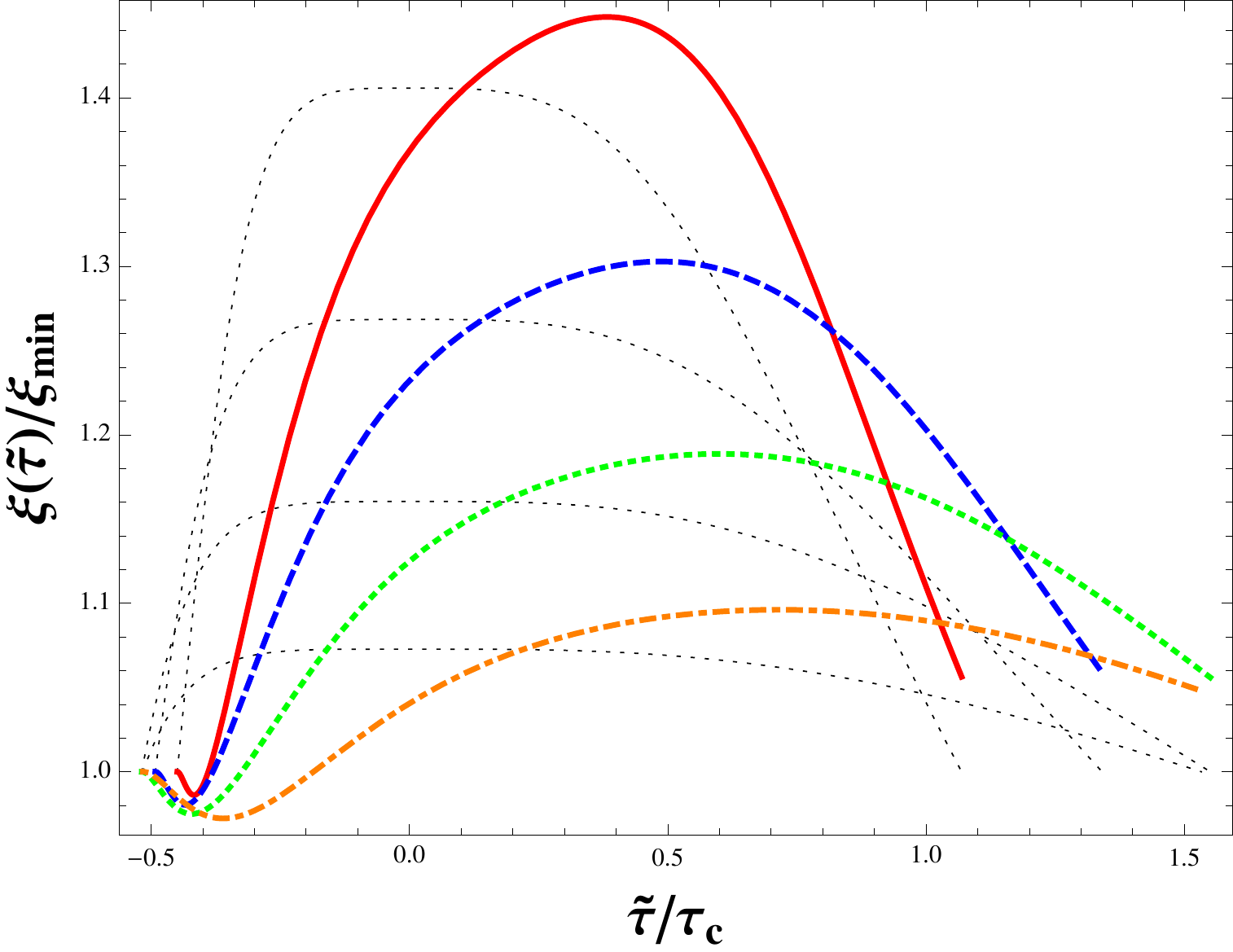}
\label{fig:kappaplotB}
}
\subfigure[]
{
\includegraphics[width=.4\textwidth]{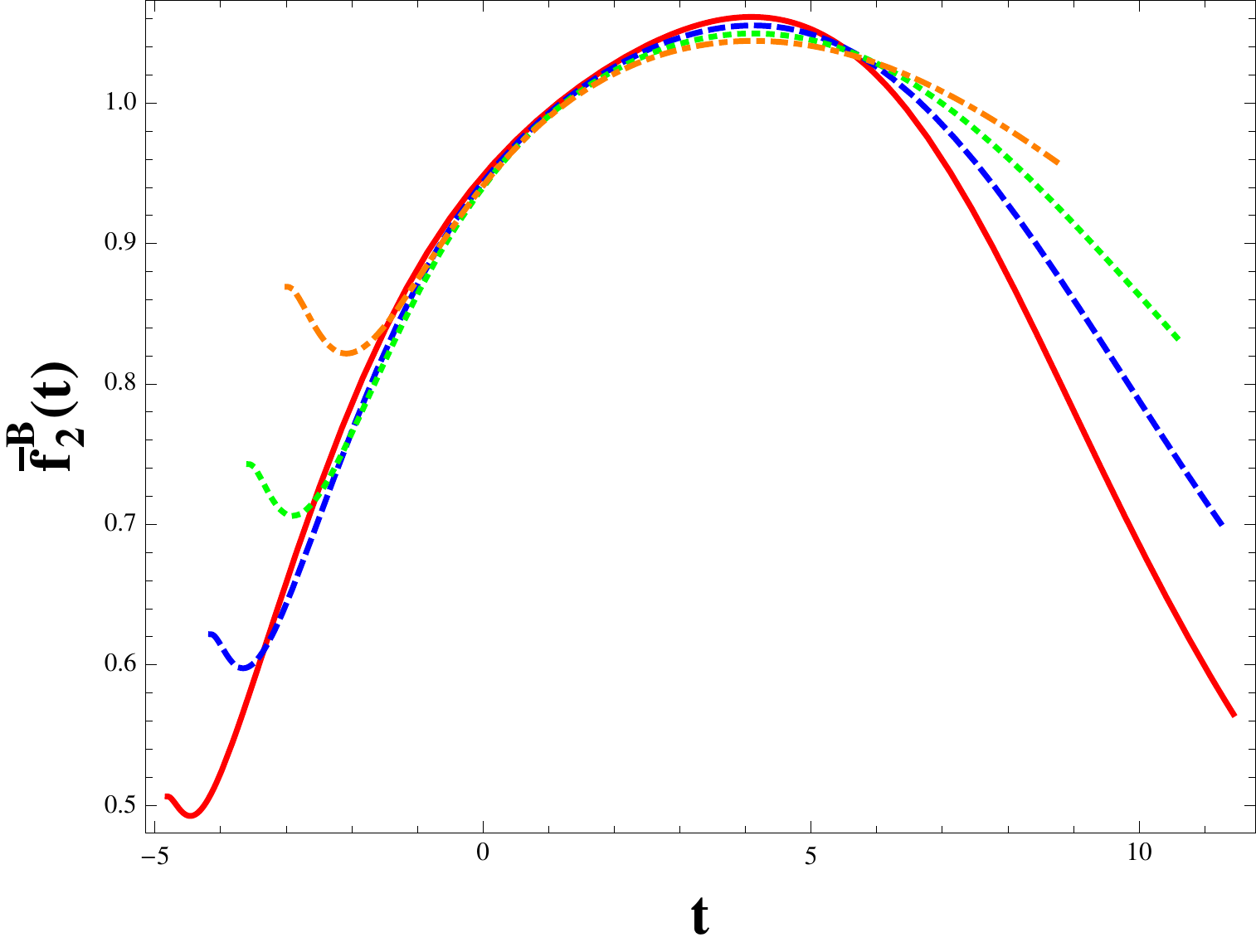}
\label{fig:f2plotB}
}
\caption{
\label{fig:secondB}
Evolution of $\xi$ (a) and ${\bar f}^{B}_2$ (b) for protocol B trajectories along the lines described in the caption for Fig.~\ref{fig:thirdB}.
 }
\end{figure}

We now present further detailed numerical tests of the non-equilibrium scaling hypothesis. 
To solve evolution equation Eq.~(\ref{eq:kappa_evo}) along a trajectory on the crossover side of the critical regime,
we need to specify 1) the trajectory in Ising phase diagram, 2) the mapping between the Ising variables $r$-$h$ and the QCD variables $T$-$\mu_{B}$, 
3) the evolution of QCD variables along the trajectory. 

Throughout this work, we will consider trajectories in which $r$ and $h$ are related by
\begin{equation}
\label{traj-para}
r= r_{c}-a_{h} h^{2}\, , 
\end{equation}
where $r_{c}$ is the value of $r$ on the cross-over line. 
As we shall see later, by changing $r_{c}$ and  $a_{h}$, we will obtain trajectories which lie in protocol A or protocol B. 
As mentioned previously, we will use the linear map $(T- T_c)/\Delta T=h$ and $(\mu_{B}-\mu^{c}_B)= -r$. 
We will also employ a simple model of the medium that mimics the expanding fireball formed in heavy ion collisions.
Specifically, we consider the evolution of temperature to be of the form
\begin{equation}
T(\tau)= T_{c}\[\frac{\tau}{\tau_{c}}\]^{-3c^{2}_{s}}\, , 
\end{equation}
and we will use $c^{2}_{s}=0.1$. 

To confirm the scaling hypothesis numerically, we first consider a representative trajectory in protocol A. 
In particular, we will consider a trajectory with fixed $r$: $a_{h}=0$ and thus $r=r_{c}$ in Eq.~(\ref{traj-para}). From the definition of protocol A, 
this trajectory will pass the crossover line in the vicinity of the critical point. Therefore $r_{c}\ll 1$. 
We will present below numerical results with $r_{c}=0.02$  in Figs.~\ref{fig:firstA}, \ref{fig:secondA}, \ref{fig:thirdA}, \ref{fig:fourthA}. 
They correspond to solutions with $\tau_{\rm rel}/\tau_{c}=0.02,0.06,0.1,0.14$. 
We have also verified the scaling behavior for other choices of $r_{c}\ll 1$. 
In producing Fig.~\ref{fig:xiplotA}, we have defined the non-equilibrium correlation length as $\xi\equiv \sqrt{\kappa_{2} V_{c}/T_{c}}$. 

We now turn to protocol B. The four trajectories representing this protocol in  Fig.~\ref{fig:trajThetaplot}  correspond to $r_{c}=0.9,0.8,0.7,0.6$ (from left to right). We fix $a_{h}$ in Eq.~(\ref{traj-para}) such that the trajectories approach the equal-$\xi_{\rm eq}$ contour in the vicinity of the cross-over line. 
This reflects the character of protocol B that the quench of the equilibrium correlation length $\xi_{\rm eq}$ is very slow near the crossover line. 
The evolution equations were solved numerically along these trajectories. To test the scaling hypothesis, 
we  tuned $\tau_{\rm rel}$ to ensure $\theta_{\rm KZ}=-0.1$ for all these trajectories. 
The prediction based on the scaling hypothesis is that the rescaled functions $f_{1,2,3,4}(\tau/\tau_{KZ})$ are independent of the choice of trajectories. 
Figs.~\ref{fig:firstB},~\ref{fig:secondB},~\ref{fig:thirdB}, ~\ref{fig:fourthB} demonstrate that there is indeed a large time window around crossover line where the scaling hypothesis works. 

\end{appendix}

\bibliography{KZ}

\end{document}